\documentclass[12pt]{article}
\usepackage[english]{babel} 
\usepackage{textcomp} 
\usepackage{amsmath,eucal,bbm,bm,mathtools} 
\usepackage{enumerate} 
\usepackage{array,multirow,rotating,longtable,arydshln,tabu} 
\usepackage{color,xcolor,bm,epstopdf,pgfplots,graphicx} 
\usepackage{caption,subcaption,float,adjustbox} 
\usepackage{algorithm} 
\usepackage[noend]{algpseudocode} 
\usepackage{eurosym}
\newcommand{\rhob}{\boldsymbol{\rho}}
\newcommand{\nub}{\boldsymbol{\nu}}
\usepackage{url} 
\usepackage[utf8]{inputenc} 
\usepackage{booktabs} 
\usepackage{microtype} 
\usepackage{makecell}
\usepackage{breqn} 
\usepackage{enumitem}
\usepackage{ulem}
\usepackage{listings}
\pgfplotsset{compat=1.16}
\usepackage{makecell}
\usepackage{breqn} 
\usepackage{enumitem}
\usepackage{ulem}
\usepackage{listings}
\usepackage{placeins}
\usepackage{hyperref}
\pgfplotsset{compat=1.16}

\usepackage{natbib}
\usepackage{url} 

\newcommand{\blind}{0}

\usepackage{xr}
\makeatletter
\newcommand*{\addFileDependency}[1]{
  \typeout{(#1)}
  \@addtofilelist{#1}
  \IfFileExists{#1}{}{\typeout{No file #1.}}
}
\makeatother

\newcommand*{\myexternaldocument}[1]{
    \externaldocument{#1}
    \addFileDependency{#1.tex}
    \addFileDependency{#1.aux}
}

\myexternaldocument{ConcProbSuppApp}

\begin{document}

\def\spacingset#1{\renewcommand{\baselinestretch}%
{#1}\small\normalsize} \spacingset{1}


\if0\blind
{
  \title{\bf Computational Efficient Approximations of the Concordance Probability in a Big Data Setting}
  \author{Robin Van Oirbeek \\
     Data Office, Allianz Benelux\\
    and \\
    Jolien Ponnet \\
	   Department of Mathematics, Section of Statistics and Data Science, KU Leuven\\
	   and \\
	   Tim Verdonck\thanks{
	   This work was supported by the the Allianz Research Chair at KU Leuven; and the Internal Funds KU Leuven under Grant C16/15/068.} \hspace{.2cm}\\
	   Department of Mathematics, Section of Applied Mathematics, UAntwerp\\
       Department of Mathematics, Section of Statistics and Data Science, KU Leuven\\
       	    \href{mailto:tim.verdonck@uantwerpen.be}{tim.verdonck@uantwerpen.be}; \href{https://orcid.org/0000-0003-1105-2028}{https://orcid.org/0000-0003-1105-2028}}
  \maketitle
} \fi

\if1\blind
{
  \bigskip
  \bigskip
  \bigskip
  \begin{center}
    {\LARGE\bf Computational Efficient Approximations of the Concordance Probability in a Big Data Setting}
\end{center}
  \medskip
} \fi

\bigskip
\begin{abstract}
Performance measurement is an essential task once a statistical model is created. The Area Under the receiving operating characteristics Curve (AUC) is the most popular measure for evaluating the quality of a binary classifier. In this case, AUC is equal to the concordance probability, a frequently used measure to evaluate the discriminatory power of the model. Contrary to AUC, the concordance probability can also be extended to the situation with a continuous response variable. Due to the staggering size of data sets nowadays, determining this discriminatory measure requires a tremendous amount of costly computations and is hence immensely time consuming, certainly in case of a continuous response variable. Therefore, we propose two estimation methods that calculate the concordance probability in a fast and accurate way and that can be applied to both the discrete and continuous setting. Extensive simulation studies show the excellent performance and fast computing times of both estimators. Finally, experiments on two real-life data sets confirm the conclusions of the artificial simulations.
\end{abstract}

\noindent%
\vfill

\newpage
\spacingset{1.5} 
\section{Introduction}
Since the beginning of this century, technological advances dramatically changed the size of data sets as well as the speed with which these data have to be processed, analyzed and evaluated. Modern data may have a huge number of observations and/or a very large number of dimensions. 
Recently, data is growing at an explosive rate and these massive data sets are collected at high speed from different sources, e.g. results of sensors in an industrial process or in healthcare, spatial data from mobile devices, etc. 
This data is often used to get insights into a certain process, which are typically obtained by predicting one response variable by all the other explanatory variables. Many statistical models can be used for this prediction task that will turn data into knowledge and hence, it is of utmost importance to compare the performance of these various models. 

For evaluating the quality of a binary classifier, the focus typically lies on the discriminatory ability of the model. For more information on the different aspects of the predictive ability, such as the difference between the discriminatory ability and the calibration, we refer to \citet{steyerberg2010assessing}. The most widely used performance measure to check the discriminatory ability of a binary classifier is the Area Under the ROC Curve (AUC), see for example the work of \cite{liu2008exploratory}. Note that the construction of the Receiver Operating Characteristic (ROC) curve was suggested by \cite{bamber1975area} with electronic signal detection theory.
It is known that the AUC in case of a binary response variable equals exactly the concordance probability, also called the C-index (see for example \citep{reddy2015healthcare}). 
The concordance probability corresponds to the probability that a randomly selected subject with outcome $Y = 1$ has a higher predicted probability $\pi(\bm{X}) = \text{P}(Y = 1 | \bm{X})$ than a randomly selected subject with outcome $Y = 0$, where $\bm{X}$ corresponds to the vector of variables \cite{pencina2004overall}:
\begin{equation}\label{ch1_eq1}
\text{C} = \text{P}\Big(\pi(\bm{X}_i) > \pi(\bm{X}_j) \; | \; Y_i = 1 , \; Y_j = 0\Big).
\end{equation}
A pair of an observation with its prediction that satisfies the above condition is called a concordant pair. Hence, the concordance probability can also be defined as the probability that a randomly selected comparable pair of observations with their predictions, is a concordant pair. The concordance probability, denoted by $\text{C}$, normally ranges between 0.5 and 1, and the closer it is to 1, the better its discriminatory ability. If its value drops below 0.5, the predictions are consistently inconsistent. For a sample of size $n$, the concordance probability typically is estimated as the ratio of the number of concordant pairs $n_c$ over the number of comparable pairs $n_t$: 

\begin{equation}\label{ch1_eq2}
\widehat{\text{C}} = \frac{n_c}{n_t} = \frac{\widehat{\pi}_c}{\widehat{\pi}_c + \widehat{\pi}_d} = \frac{\sum_{i = 1}^{n - 1}\sum_{j = i + 1}^{n}I\Big(\widehat{\pi}(\bm{x}_i) > \widehat{\pi}(\bm{x}_j), \; y_i = 1, \; y_j = 0\Big)}{\sum_{i = 1}^{n - 1}\sum_{j = i + 1}^{n}I\Big(\widehat{\pi}(\bm{x}_i) \neq \widehat{\pi}(\bm{x}_j), \; y_i = 1, \; y_j = 0\Big)}, 
\end{equation}

where the value $\widehat{\pi}_c$ (respectively $\widehat{\pi}_d$) refers to the estimated probability that a comparable pair is concordant (respectively discordant) and $I(\cdot)$ to the indicator function. For observation $i$, $y_i$ and $\widehat{\pi}(\bm{x}_i)$ correspond to the observed outcome and estimated predicted probability with $\bm{X}_i$ the vector of observed variables respectively. Note that the extra condition $\widehat{\pi}(\bm{x}_i) \neq \widehat{\pi}(\bm{x}_j)$ is added to the denominator to ensure that no ties in the predictions are taken into account \citep{yan2008investigating}. Ties namely attenuate the concordance probability to 0.5, depending on the distribution of the observations over the different risk groups, which is undesirable \citep{heller2016estimating}. According to \cite{yan2008investigating}, it depends on the context whether one wants to ignore ties. An estimator should therefore be able to handle both situations. Since ties only influence the actual definition of the comparable and concordant pairs, the choice of whether to include ties does not influence the expression of any definition or estimator presented here. See \cite{yan2008investigating} for a more detailed treatment of the subject of ties on the concordance probability.

As can be seen in Equation \eqref{ch1_eq2}, calculating this C-index can be very time consuming for large data sets. Therefore, research was conducted to find approximations that reduce the time complexity for calculating such a discriminatory measure in a discrete setting. For example, \cite{bouckaert2006efficient} provides a method for incrementally updating exact AUC curves and for calculating approximate AUC curves. A number of researchers have attempted to approximate the AUC by smooth functions \citep{komori2011boosting,eguchi2002class,ma2005regularized}, or more specifically polynomials \citep{calders2007efficient}. The most popular algorithm is introduced by \citet{fawcett2006introduction}, where the ROC curve is determined efficiently such that the AUC can be calculated as the area under the ROC curve. This area is obtained by approximating the integral by the trapezium rule.

When the response variable $Y$ is continuous, the discriminatory ability of the model can also be of interest, especially when the ranking of the observations is of higher importance than obtaining well-calibrated predictions. Typical examples are the modelling of the claim size distribution of a non-life insurance product, or the modelling of a lifetime distribution in a churn analysis. Note that the concordance probability is particularly popular within the field of survival analysis, and many estimation methods have been proposed to compute the concordance probability in the presence of right censoring \citep{Harrell1982, Gerds2007}. As such, the basic Definition \eqref{ch1_eq1} is typically adapted as:
\begin{equation}\label{ch1_cont_eq1}
\text{C} = \text{P}\Big(\pi(\bm{X}_i) > \pi(\bm{X}_j) \; | \; Y_i > Y_j \Big),
\end{equation}
where $\pi(\bm{X}_i)$ represents the predicted value of $Y_i$. 
However, in some situations distinguishing nearly identical observations has little practical importance. Consequently, the basic definition in \eqref{ch1_cont_eq1} can be extended as follows:
\begin{equation}\label{ch1_cont_eq2}
\text{C} = \text{P}\Big(\pi(\bm{X}_i) > \pi(\bm{X}_j) \; | \; Y_i - Y_j > \nu  \Big),
\end{equation}
where $\nu\geq 0$. In other words, the difference between the considered observations is at least $\nu$. Note that in this continuous setting, there is no link with the area under the ROC curve. As a result, the aforementioned research is not suitable for this setting in  case of a large dataset. 

In this paper, we present two computational efficient approximations of the concordance probability for big data, both in a discrete and continuous setting. The first one is based on the $k$-means clustering algorithm, whereas the second one makes use of an interesting structure that can be found in the estimation process of the concordance probability. The remainder of this paper is structured as follows. Section \ref{sec: approx} presents a detailed explanation of the two proposed approximations. The good performance of both estimators of the concordance probability is verified in an extensive simulation study in Section \ref{sec: sim}. Finally, we illustrate the practical applications of this work on two real-life examples in Section \ref{sec: ex}. The main findings and suggestions for further research are summarized in the conclusion.

\section{\textit{{k}}-means and marginal approximation}
\label{sec: approx}
In this section, two approximations of the concordance probability are presented. One approximation makes use of the $k$-means clustering algorithm and will therefore be called the \textit{$k$-means approximation}. The other one takes advantage of the structure of the estimation process of the concordance probability and will be denoted the \textit{marginal approximation}. The first subsection focuses on the discrete setting, meaning that the response variable is binary. The second subsection considers the continuous setting, where the response variable is continuous.
\subsection{Discrete setting}
\label{subsec: approx discrete}
In this subsection, we consider the situation of a binary response variable $Y$. The $k$-means approximation can be obtained by separately applying a clustering algorithm to the predictions of each group in the definition of the concordance probability. The number of clusters $k$ will then determine the level of approximation, hence a more precise estimate will be obtained as $k$ increases. When the clustering algorithm is applied, only the cluster centroids will be considered for estimating the concordance probability. In other words, all clusters of each group will be compared with one another. The importance of each cluster by cluster comparison is weighted by the probability that a randomly selected pair of observations belongs to the respective clusters, or:
\begin{align}
\label{ch1_eq8}
\widehat{\pi}_c &\approx \sum_{l = 1}^{k}\sum_{m = 1}^{k}I\Big(\widehat{\pi}_B^l > \widehat{\pi}_A^m\Big)w_B^lw_A^m, \nonumber \\
\widehat{\pi}_d &\approx \sum_{l = 1}^{k}\sum_{m = 1}^{k}I\Big(\widehat{\pi}_B^l < \widehat{\pi}_A^m\Big)w_B^lw_A^m, \nonumber \\
\widehat{\text{C}}_{\text{$k$-means}} &= \frac{\widehat{\pi}_c}{\widehat{\pi}_c + \widehat{\pi}_d}\approx \frac{\sum_{l = 1}^{k}\sum_{m = 1}^{k}I\Big(\widehat{\pi}_B^l > \widehat{\pi}_A^m\Big)w_B^lw_A^m}{\sum_{l = 1}^{k}\sum_{m = 1}^{k}I\Big(\widehat{\pi}_B^l \neq \widehat{\pi}_A^m\Big)w_B^lw_A^m},
\end{align}

where subscript B (respectively A) refers to the group of observations of which the predictions are supposed to be higher (respectively lower) than the ones of subscript A (respectively B). The denominator of the estimator is needed to eliminate the effect of ties on the predictions. Moreover, $\widehat{\pi}_*^l$ is the centroid of the $l$-th cluster of group * and $w_*^l$ the weight of the $l$-th cluster of group *. The latter is estimated by computing the percentage of observations of group * that belongs to cluster $l$, such that by definition $\sum_{l = 1}^{k}w_*^l = 1$. 
\newline

The marginal approximation is obtained by taking advantage of a structure that can be found in the estimation process of the concordance probability. For each observation of a given group, its prediction is compared to all predictions of the other group. As such, no correlation is present between the pairs $(\pi_A, \pi_B)$, yielding the following expression for the bivariate distribution $F_{\pi_A,\pi_B}(\pi_A, \pi_B)$:

\begin{equation*}
F_{\pi_A,\pi_B}(\pi_A, \pi_B) = F_{\pi_A}(\pi_A)F_{\pi_B}(\pi_B)
\end{equation*}

\begin{figure}[tb]
\centering
\includegraphics[width=8cm]{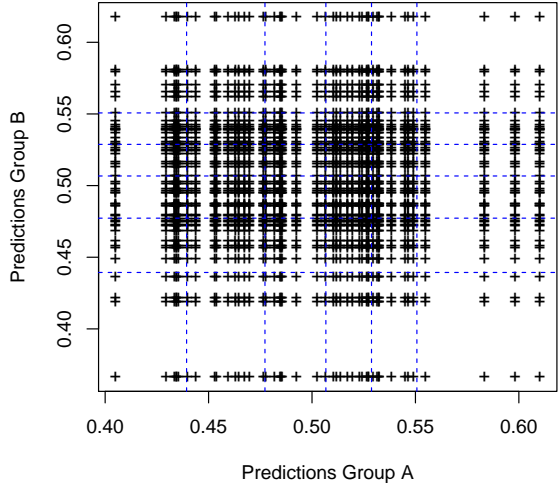}
\caption{\label{fig1} Scatterplot showing the different pairs that are considered in the estimation of the concordance probability. In dashed blue lines the grid lines are shown, hereby delineating the different regions of the grid.}
\end{figure}

Hence, when a grid with the same $q$ boundary values $\bm{\tau} = (\tau_0 \equiv -\infty, \tau_1, \ldots, \tau_q, \tau_{q + 1} \equiv +\infty)$ for the marginal distribution of both groups is placed on top of the latter bivariate distribution, the probability that a pair belongs to any of the delineated regions only depends on the marginal distributions $F_{\pi_A}(\pi_A)$ and $F_{\pi_B}(\pi_B)$. This idea is depicted in Figure \ref{fig1}.

\begin{figure}[tb]
\centering
\includegraphics[width=8cm]{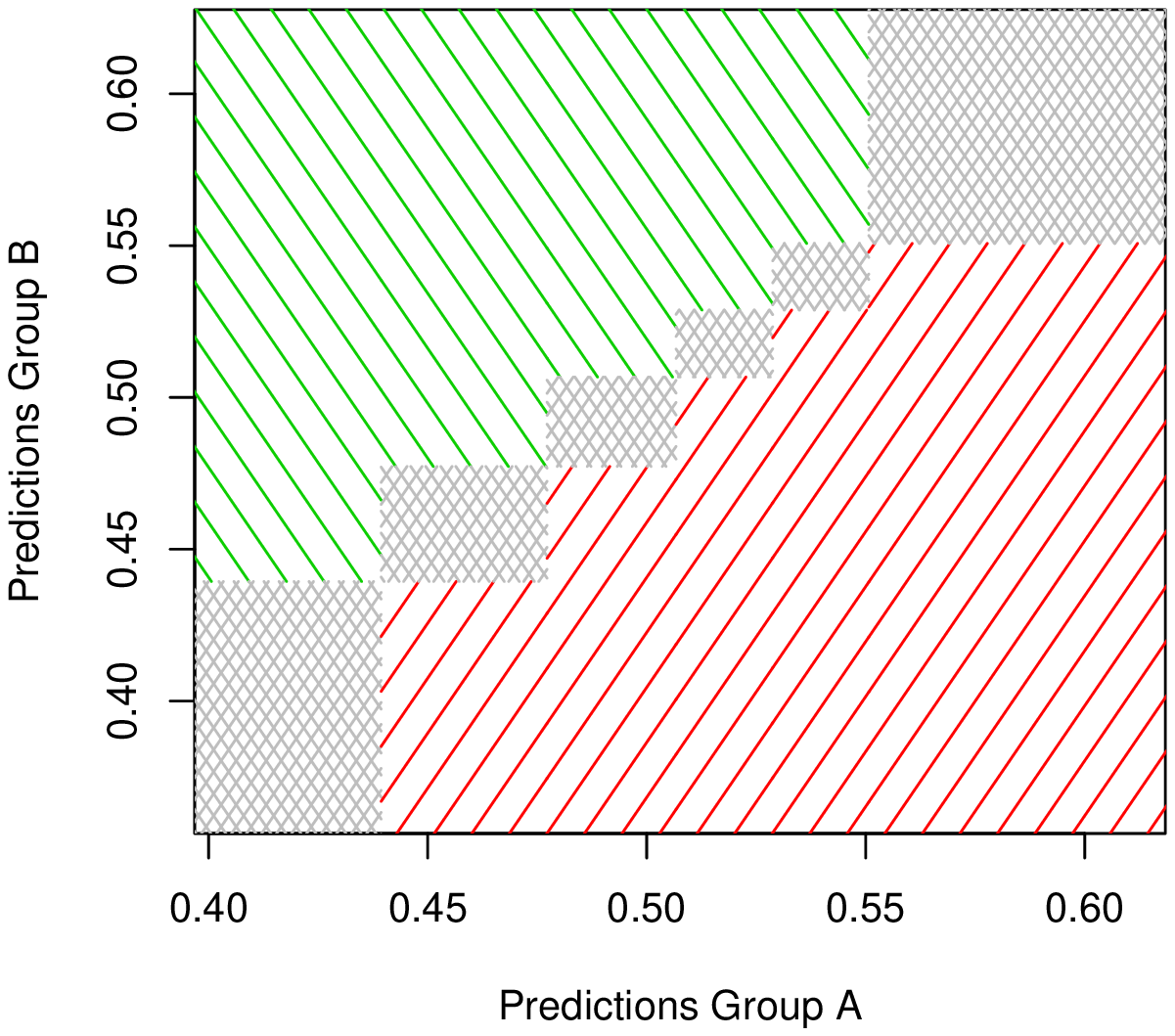}
\caption{\label{fig2} The different regions of the grid, shown in Figure \ref{fig1}, in which the concordant pairs (downward dashed region, in green), the discordant pairs (upward dashed region, in red) and incomparable pairs (upward and downward dashed region, in grey) are highlighted.}
\end{figure}

In a next step, three different regions in the two dimensional space can be determined: regions
that only contain concordant pairs, regions that contain only discordant pairs, and regions that also contain ties (induced by the grid structure). As such the latter region is considered to be a region of incomparable pairs only. For the ease of presentation, we assume that the $q$ boundaries are the same for both groups, but this is not an absolute necessity.  Remember that it is assumed that all observations of group A have a lower observed outcome value than all the observations of group B, and that the values of group A (respectively B) are plotted on the X (respectively Y)-axis. Therefore, all concordant pairs are located in the upper left part of the support of $F_{\pi_A,\pi_B}(\pi_A, \pi_B)$, whereas all discordant pairs are in the lower right part of the support of $F_{\pi_A,\pi_B}(\pi_A, \pi_B)$, and the incomparable pairs are in between both (see Figure \ref{fig2} for a visual representation).

Since only both marginal distributions $F_{\pi_A}(\pi_A)$ and $F_{\pi_B}(\pi_B)$ are needed to compute the probability that a pair belongs to a region of the support of $F_{\pi_A,\pi_B}(\pi_A, \pi_B)$, the concordance probability can be estimated as follows:
\begingroup
\allowdisplaybreaks
\begin{align*}
\displaystyle \widehat{\pi}_c &\approx \sum_{i = 1}^q \Big(\widehat{F}_{\pi_A}(\tau_i) - \widehat{F}_{\pi_A}(\tau_{i - 1})\Big)\Big(1 - F_{\pi_B}(\tau_i)\Big),\nonumber\\
\widehat{\pi}_d &\approx \sum_{i = 2}^{q + 1} \Big(\widehat{F}_{\pi_A}(\tau_i) - \widehat{F}_{\pi_A}(\tau_{i - 1})\Big)F_{\pi_B}(\tau_{i - 1}),\nonumber \\
\widehat{\text{C}}_{\text{marg}}& = \frac{\widehat{\pi}_c}{\widehat{\pi}_c + \widehat{\pi}_d} \\
&\approx \frac{\sum\limits_{i = 1}^q \Big(\widehat{F}_{\pi_A}(\tau_i) - \widehat{F}_{\pi_A}(\tau_{i - 1})\Big)\Big(1 - F_{\pi_B}(\tau_i)\Big)}{\sum\limits_{i = 1}^{q + 1} \Big(\widehat{F}_{\pi_A}(\tau_i) - \widehat{F}_{\pi_A}(\tau_{i - 1})\Big)\Big[I(i \neq q + 1)\Big(1 - F_{\pi_B}(\tau_i)\Big) + I(i \neq 1)F_{\pi_B}(\tau_{i - 1})\Big]}.
\end{align*}
\endgroup
Note that this methodology only yields an approximation of the estimated concordance probability of \eqref{ch1_eq2}, and that the accuracy improves as $q$ increases. Although $\bm{\tau}$ can be determined in many ways, we strongly recommend to base it on a set of evenly spaced percentiles of the empirical distribution of the predictions of both group A and B jointly. This is motivated by the easiness of determining the number of observation-prediction couples in each grid.

\subsection{Continuous setting}
\label{subsec: approx cont}
In this subsection, we consider the situation of a continuous response variable $Y$. Since the data structure and the design of the concordance probability is entirely different for the continuous setting as compared to the discrete setting, the approximations that were proposed in Section \ref{subsec: approx discrete} will not necessarily work for this new setting. One of the key points underpinning these approximations is the existence of two independent groups. In case of a continuous response variable, two of such groups cannot be defined and therefore the previous approximations need to be adapted to the continuous setting.

Based on \eqref{ch1_cont_eq2},  the large data set can only be reduced to a smaller set of clusters, when these clusters of observations are jointly constructed based on their observed outcomes and predictions. All these clusters are uniquely characterized by their observed outcome, prediction and weight, the latter being determined by the number of observations that pertain to the cluster at hand. As a result, \eqref{ch1_cont_eq2} can be computed whilst using only these representations of the clusters,

\begin{equation*}
\widehat{\pi}_c \approx \sum_{i = 1}^{k - 1}\sum_{j = i + 1}^{k}I\Big(\widehat{\pi}^i > \widehat{\pi}^j, \; y^{i} - y^{j} > \nu\Big)w^iw^j,
\end{equation*}

\begin{equation*}
\widehat{\pi}_d \approx \sum_{i = 1}^{k - 1}\sum_{j = i + 1}^{k}I\Big(\widehat{\pi}^i < \widehat{\pi}^j, \;  y^{i} - y^{j} > \nu\Big)w^iw^j,
\end{equation*}

\begin{equation}\label{ch1_eq11}
\widehat{\text{C}}_{\text{$k$-means}}(\nu) = \frac{\widehat{\pi}_c}{\widehat{\pi}_c + \widehat{\pi}_d} \approx \frac{\sum_{i = 1}^{k - 1}\sum_{j = i + 1}^{k}I\Big(\widehat{\pi}^i > \widehat{\pi}^j, \; y^{i} - y^{j} > \nu\Big)w^iw^j}{\sum_{i = 1}^{k - 1}\sum_{j = i + 1}^{k}I\Big( y^{i} - y^{j} > \nu\Big)w^iw^j},
\end{equation}

where $y^{l}$ and $\widehat{\pi}^l$ are the observed outcome and the prediction of the representation of the $l$-th cluster respectively; which is the centroid in case of $k$-means. $w^l$ is the weight of the $l$-th cluster that is estimated by computing the percentage of observations belonging to cluster $l$, such that by definition $\sum_{l = 1}^{k}w^l = 1$. Note that the number of clusters, necessary to obtain a good approximation, is likely much higher for this definition than for Definition \eqref{ch1_eq8} since Definition \eqref{ch1_eq11} needs to consider the additional condition $y^{i} - y^{j} > \nu$, which is not needed in the discrete setting. Clearly, in order to ensure that a sufficient number of clusters are comparable for larger values of $\nu$, $k$ should be large enough. 

The marginal approximation can be adapted to the continuous setting as well. In this case, a grid will be placed on the $(Y, \pi(\bm{X}))$ space. Since the condition in Definition \eqref{ch1_cont_eq2} only takes the continuous variable $Y$ into account, the $q$ boundary values $\bm{\tau} = (\tau_0 \equiv -\infty, \tau_1, \ldots, \tau_q, \tau_{q + 1} \equiv +\infty)$ are based on the observed values of $Y$ only. Similarly as for the discrete setting, the set $\bm{\tau}$ will be selected as a set of evenly spaced percentiles from the empirical distribution of the observed values for $Y$. The set of boundary values for dimension $\pi(\bm{X})$ do not necessarily need to be the same as the one for dimension $Y$. However, we will use the same boundary values nevertheless, as this will allow for better visualization of the discriminatory ability of the model over the $(Y, \pi(\bm{X}))$, as will be apparent below.

An important difference with the marginal approximation in the discrete setting is that the distance in the $Y$ dimension between two regions needs to be larger than $\nu$ before they can be compared. Definition \eqref{ch1_cont_eq2} states namely that the difference in observed values of $Y$ should be at least $\nu$. Therefore, regions of which the distance in the $Y$ dimension between their lower boundary and the upper boundary of the selected region is smaller than $\nu$, should not be considered when estimating the selected region's contribution to $\widehat{\text{C}}(\nu)$. These eliminated regions potentially contain observations of which its observed value of $Y$ does not differ at least $\nu$ from every observation of the selected region. 

\begin{figure}[h!]
\centering
\includegraphics[width=8cm]{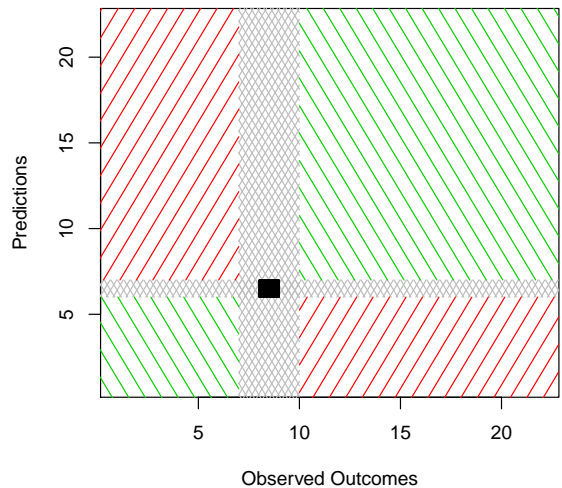}
\caption{\label{fig3} A grid is laid on top of the $(Y, \pi)$ space in which the regions are highlighted that contain the concordant pairs (downward dashed region, in green), the discordant paired (upward dashed region in red) and incomparable pairs (upward and downward dashed region, in grey) for the region shown in an even black rectangle.}
\end{figure}

Ties in the predictions need to be eliminated as well, such that regions that have exactly the same boundary values for the $\pi(\bm{X})$ dimension as the selected region, will not be considered for the estimation of $\widehat{\text{C}}(\nu)$ either. The effect of eliminating both sets of regions, is that the $(Y, \pi(\bm{X}))$ space is partitioned in two or four main regions, depending on whether the selected region is located on the border of the grid. This partitioning greatly simplifies the estimation of the contribution of the selected region, since for each of the two or four main regions, all members of the selected region either contribute to $\widehat{\pi}_c$ or $\widehat{\pi}_d$. Interestingly is that in all cases, the number of main regions that contribute to $\widehat{\pi}_c$ is always equal to the number of main regions that contribute to $\widehat{\pi}_d$. A visual representation of the above effect is shown in Figure \ref{fig3}.

During the estimation of $\widehat{\text{C}}_{\text{marg}}(\nu)$, it is important to let each region-to-region comparison contribute only once. This can be obtained by only considering those main regions during the computation of the total contribution of a selected region, that are located in the same, fixed direction as compared to the selected region. For example, by only considering those main regions that are located on the right-hand side of the selected region in the $(Y, \pi(\bm{X}))$ space, i.e. those regions that have a higher observed value for $Y$ than the one of the selected region, each region-to-region comparison will only contribute once. One might as well choose any of the other three remaining directions, i.e. on the left-hand side, below or above the selected region, as long as the same direction is used for all selected regions. Therefore, by selecting the main regions that are located on the right-hand side of the selected region only, the marginal approximation of the concordance probability can be computed as:
\allowdisplaybreaks 
\begin{align*}
n_C(\nu) &\approx \sum_{i = 1}^{q+1}\sum_{j = 1}^{q+1} \Big(n_{C,\tau_{ij}}^\rightarrow(\nu)\Big) \nonumber\\
&= \sum_{i = 1}^{q+1}\sum_{j = 1}^{q+1} \Bigg( I\Big(i \le q, j \le q \Big) \sum_{k = i + 2}^{q+1}\sum_{l = j + 1}^{q+1} \Bigg( I\Big(\tau_i +\nu \le \tau_{k-1} \Big)n_{\tau_{ij}, \tau_{kl}} \Bigg), \nonumber\\
n_D(\nu) &\approx \sum_{i = 1}^{q+1}\sum_{j = 1}^{q+1} \Big(n_{D,\tau_{ij}}^\rightarrow(\nu)\Big)\nonumber\\
&= \sum_{i = 1}^{q+1}\sum_{j = 1}^{q+1}  \Bigg( I\Big(i \le q, j \ge 2 \Big)\sum_{k = i + 1}^{q+1}\sum_{l = 1}^{j-1} \Bigg( I\Big(\tau_i +\nu \le \tau_{k-1} \Big)n_{\tau_{ij}, \tau_{kl}} \Bigg),\nonumber\\
\widehat{\text{C}}_{\text{marg}}(\nu)& = \frac{n_C(\nu)}{n_C(\nu) + n_D(\nu)} \approx \widehat{\text{C}}_{\text{marg}}^\rightarrow(\nu) = \frac{\sum_{i = 1}^{q + 1}\sum_{j = 1}^{q + 1} \Big(n_{C,\tau_{ij}}^\rightarrow(\nu)\Big)}{\sum_{i = 1}^{q + 1}\sum_{j = 1}^{q + 1} \Big(n_{C,\tau_{ij}}^\rightarrow(\nu) + n_{D,\tau_{ij}}^\rightarrow(\nu)\Big)},
\end{align*}

where, $\tau_{ij}$ corresponds to the rectangle with values $Y \in \; [\tau_{i - 1}, \tau_{i}[$ and values $\pi(\bm{X}) \in \; [\tau_{j - 1}, \tau_{j}[$. Furthermore, $n_{C,\tau_{ij}}^\rightarrow(\nu)$ ($n_{D,\tau_{ij}}^\rightarrow(\nu)$) equals the number of concordant (discordant) comparisons for region $\tau_{ij}$, and $n_{\tau_{ij}, \tau_{kl}}$ is the product of the number of elements in regions $\tau_{ij}$ and $\tau_{kl}$.

\FloatBarrier
\section{Simulations}
\label{sec: sim}
We now investigate the performance of the proposed $k$-means approximation and the marginal approximation for the concordance probability in an extensive simulation study. Section \ref{sec: disc sim} investigates the approximations for the discrete setting with a binary response variable, whereas Section \ref{sec: cont sim} focuses on the continuous setting with a continuous response variable.
\subsection{Discrete setting}
\label{sec: disc sim}
\textbf{Data generation setup} 
 To examine the performance of the proposed estimators when the response variable is binary, a simulation study was set up on a response variable following a beta-binomial distribution with parameter $n$ equal to 1.
The beta-binomial distribution, denoted $BB(\alpha,\beta,n)$ with $\alpha>0$ and $\beta>0$, is a compound probability distribution and can be seen as a binomial distribution where the parameter $p$  (i.e. probability of success) is randomly drawn from a beta distribution, denoted $\text{Beta}(\alpha,\beta)$. Note that when $Y \sim \text{Beta}(\alpha, \beta)$, its mean $\mu$ equals $\frac{\alpha}{\alpha+\beta}$ and its concentration is defined by $\kappa=\alpha+\beta$. The concentration indicates how broad this beta distribution is: the larger the concentration, the narrower the distribution. Moreover, we focus on $\alpha>1$ and $\beta$>1 since the beta distribution is then unimodal. More information about the beta-binomial distribution can be found in Appendix A.

In this simulation setting, a parameter $p$ is sampled from the beta distribution (and used for true prediction), whereas a sample of the corresponding beta-binomial distribution yields the observed value. These pairs of predicted and observed values for $p$ are then used to calculate the proposed approximations for the concordance probability.

	
First, the population value of $C$ is computed. For this, a large sample (i.e. with sample size 100,000,000) is selected from a beta distribution. Based on its mean and concentration, we considered 63 possible beta distributions (resulting in 63 population values), see Table \ref{tab: popVal} for combinations. Next, these samples are used as true rate values to sample outcomes from the binomial distribution with $n=1$ (i.e. Bernouilli distribution). These sampled outcomes, i.e. the sampled rates, can be seen as the true predictions coming from the beta distribution. Moreover, they can be classified into two groups: the ones that have a sampled outcome equal to 0 and the ones that have a sampled outcome equal to 1. The population value is then computed by comparing the sampled rates of the 0 group with the ones of the 1 group. This population value is for each of the considered situations presented in Table \ref{tab: popVal}. 

A first thing to notice is that a concentration $\kappa=15$ was not possible for extreme probabilities $p$, since it required $\alpha$ or $\beta$ to be smaller than 1. As expected, we do see similarities in \textit{complementary} probabilities; e.g. $p=0.25$ has approximately the same population values as $p=0.75$. In these situations, the relative size of the two groups remains the same. Moreover, the larger the concentration, the smaller the population value of the concordance probability. This can be explained by the fact that the difference between the values of the probabilities, i.e. the predictions, is small in case of a high concentration. Finally, the more extreme the probability (close to zero or close to one), the higher the concordance probability.
\begin{table}[ht]
\centering
\scriptsize
\caption{The population values of $C$ when $p$ is sampled from different beta distributions with mean $\mu$ and concentration $\kappa$.}
\label{tab: popVal}
\begin{tabular}{ccccccccc}
   \hline \hline
&\multicolumn{8}{c}{$\mu$}\\
  \cline{3-9}
$\kappa$ && $0.05$ & $0.10$ & $0.20$ & $0.25$ & $0.30$ & $0.40$ & $0.50$   \\ 
  \hline
  15 &&& 0.7234 & 0.6744 & 0.6623 & 0.6541 & 0.6448 & 0.6421  \\ 
  25 && 0.7347 & 0.6788 & 0.6374 & 0.6276 & 0.6208 & 0.6134 & 0.6111 \\
 50 && 0.6741 & 0.6297 & 0.5985 & 0.5912 & 0.5862 & 0.5808 & 0.5792  \\ 
  100 && 0.6263 & 0.5929 & 0.5701 & 0.5648 & 0.5613 & 0.5574 & 0.5562  \\ 
  150 && 0.6041 & 0.5762 & 0.5574 & 0.5529 & 0.5501 & 0.5469 & 0.5460 \\ 
   \multicolumn{9}{c}{}\\
   &\multicolumn{8}{c}{$\mu$}\\
   \cline{3-9}
    $\kappa$&& $0.60$ & $0.70$ & $0.75$ & $0.80$ & $0.90$ & $0.95$ \\ 
  \hline
    15 &&   0.6448 & 0.6541 & 0.6623 & 0.6744 & 0.7234 &  \\ 
    25 &&  0.6134 & 0.6208 & 0.6275 & 0.6374 & 0.6788 & 0.7347\\
  50 && 0.5808 & 0.5862 & 0.5911 & 0.5985 & 0.6297 & 0.6741 \\ 
  100 &&  0.5574 & 0.5613 & 0.5648 & 0.5701 & 0.5929 & 0.6263 \\ 
  150 && 0.5469 & 0.5501 & 0.5530 & 0.5574 & 0.5762 & 0.6041 \\ 
     \hline \hline
\end{tabular}
\end{table}

It is important to investigate whether the size of the population value has an effect on the algorithms. Therefore, we  consider $\mu = 0.10$, for $\kappa \in \{15, 50, 150\}$ in the above setting. To determine whether the extremity of the probability affects the algorithms, the simulation study also focuses on $\mu \in \{0.10, 0.25, 0.50\}$, for $\kappa = 50$. Lastly, the final part of the simulation study focuses on $\mu \in \{0.05, 0.25, 0.75, 0.95\}$ for $\kappa=50$, to investigate whether only the relative size between the two groups matters or whether it is important which group is the smallest one.

\textbf{Evaluation setup}
For each of the above discussed simulation settings, 1,000 samples are generated on which the $k$-means approximation and the marginal approximation are applied to calculate the concordance probability. As benchmark method, we added the standard trapezium approximation of \citet{fawcett2006introduction}, which determines the ROC curve efficiently and then calculates the AUC as the area under the ROC curve, by approximating this integral by the trapezium rule. 

This simulation study also tests the effect of using 10, 20, 100, 500 or 1,000 boundary values or clusters, while working with a data set with 500,000 or 5,000,000 observations. In case of the marginal approximation, the boundary values are evenly spaced percentiles of the empirical distribution of the predictions of both groups, as advised in Section \ref{subsec: approx discrete}. Focusing on the simulation setting with $\mu=0.1$ and $\kappa = 50$, Table \ref{tab: mean-med, mu=0.1, conc=50} shows the bias (based on the mean or the median) together with the mean and median run time, whereas Table \ref{tab: sd-IQR, mu=0.1, conc=50} contains the standard deviation and the interquartile range of the computed concordance measure and run time. For the other simulation settings, the same tables are constructed and can be seen in Appendix B. 

\begin{table}[ht]
\centering
\scriptsize
\caption{This table considers the discrete simulation setting with  $\mu=0.10$ and $\kappa=50$. The mean and median bias and run time are shown for the approximation based on the trapezium rule, the marginal approximation (in function of the number of boundaries) and the $k$-means approximation (in function of the number of clusters). Two data set sizes are considered, namely 500,000 and 5,000,000.}
\label{tab: mean-med, mu=0.1, conc=50}
\begin{tabular}{cccccccccccc}
\hline\hline
  & \multicolumn{5}{c}{bias} & & \multicolumn{5}{c}{run time (s)}\\ 
 \cline{2-6} \cline{8-12} 
\multirow{2}{*}{\textbf{\textit{k}-means}} & \multicolumn{2}{c}{\textbf{500,000}} & & \multicolumn{2}{c}{\textbf{5,000,000}} & & \multicolumn{2}{c}{\textbf{500,000}} & & \multicolumn{2}{c}{\textbf{5,000,000}} \\ 
\cline{2-3} \cline{5-6} \cline{8-9} \cline{11-12} 
&  mean & median & & mean & median & & mean & median & & mean & median \\ \hline
10   & -0.0002 & 0.0041 &  & -0.0007 & 0.0009 &  & 0.5616 & 0.5215 &  & 4.8458 & 4.6855 \\ 
  20   & -0.0002 & -0.0001 &  & -0.0002 & -0.0004 &  & 0.5677 & 0.5270 &  & 4.5939 & 3.9200 \\ 
  100   & -0.0001 & -0.0001 &  & -0.0002 & -0.0002 &  & 0.8358 & 0.7740 &  & 6.3073 & 6.2630 \\
  500   & -0.0001 & -0.0001 &  & -0.0002 & -0.0002 &  & 4.7549 & 4.4940 &  & 19.9356 & 19.7370 \\ 
  1,000   & -0.0001 & -0.0001 &  & -0.0002 & -0.0001 &  & 14.2775 & 13.4840 &  & 42.5733 & 41.8800 \\ 
\hline
\multirow{2}{*}{\textbf{marginal}} & \multicolumn{2}{c}{\textbf{500,000}} & & \multicolumn{2}{c}{\textbf{5,000,000}} & & \multicolumn{2}{c}{\textbf{500,000}} & & \multicolumn{2}{c}{\textbf{5,000,000}} \\ 
\cline{2-3} \cline{5-6} \cline{8-9} \cline{11-12} 
 & mean & median & & mean & median & & mean & median & & mean & median \\ \hline
 10  & 0.0116 & 0.0116 &  & 0.0115 & 0.0115 &  & 0.1109 & 0.1100 &  & 1.2258 & 1.2190 \\ 
  20  & 0.0060 & 0.0060 &  & 0.0060 & 0.0060 &  & 0.1112 & 0.1110 &  & 1.2274 & 1.2220 \\ 
  100  & 0.0012 & 0.0012 &  & 0.0011 & 0.0011 &  & 0.1214 & 0.1210 &  & 1.2867 & 1.2810 \\ 
  500  & 0.0002 & 0.0001 &  & 0.0001 & 0.0001 &  & 0.1683 & 0.1670 &  & 1.6043 & 1.6060 \\ 
  1,000  & 0.0000 & 0.0000 &  &  0.0000 &  0.0000 &  & 0.2244 & 0.2230 &  & 1.9884 & 1.9870 \\ 
\hline
\multirow{2}{*}{\textbf{trapezium}} & \multicolumn{2}{c}{\textbf{500,000}} & & \multicolumn{2}{c}{\textbf{5,000,000}} & & \multicolumn{2}{c}{\textbf{500,000}} & & \multicolumn{2}{c}{\textbf{5,000,000}} \\ 
\cline{2-3} \cline{5-6} \cline{8-9} \cline{11-12} 
& mean & median & & mean & median & & mean & median & & mean & median \\ \hline
 & -0.0001 & -0.0001 &  & -0.0002 & -0.0001 &  & 0.3028 & 0.3010 &  & 3.0180 & 3.0180 \\ 
\hline\hline
\end{tabular}
\end{table}

\begin{table}[ht]
\centering
\scriptsize
\caption{This table considers the discrete simulation setting with $\mu=0.10$ and $\kappa=50$. The interquartile range and standard deviation over the estimates and over the run time are shown in function of the number of boundary values or clusters. This is done for the
$k$-means approximation, the marginal approximation and the one based on the trapezium rule. Two data set sizes are considered, namely 500,000 and 5,000,000.}
\label{tab: sd-IQR, mu=0.1, conc=50}
\begin{tabular}{ccccccccccccc}
\hline\hline
 &  \multicolumn{5}{c}{$\hat{C}$} & & \multicolumn{5}{c}{run time (s)}\\ 
 \cline{2-6} \cline{8-12} 
\multirow{2}{*}{\textbf{\textit{k}-means}}  & \multicolumn{2}{c}{\textbf{500,000}} & & \multicolumn{2}{c}{\textbf{5,000,000}} & & \multicolumn{2}{c}{\textbf{500,000}} & & \multicolumn{2}{c}{\textbf{5,000,000}} \\ 
\cline{2-3} \cline{5-6} \cline{8-9} \cline{11-12} 
& $\sigma$ & IQR & & $\sigma$ & IQR & & $\sigma$ & IQR & & $\sigma$ & IQR \\ \hline
10 &   0.0205 & 0.0252 &  & 0.0161 & 0.0216 &  & 0.1897 & 0.2463 &  & 1.5367 & 1.9117 \\ 
  20   & 0.0046 & 0.0059 &  & 0.0042 & 0.0057 &  & 0.1919 & 0.2690 &  & 1.4260 & 1.8145 \\ 
  100  & 0.0013 & 0.0017 &  & 0.0005 & 0.0007 &  & 0.1531 & 0.1063 &  & 0.1740 & 0.1490 \\ 
  500   & 0.0013 & 0.0017 &  & 0.0004 & 0.0005 &  & 0.5185 & 1.0410 &  & 0.5131 & 0.2822 \\ 
  1,000   & 0.0013 & 0.0017 &  & 0.0004 & 0.0005 &  & 1.1108 & 2.1002 &  & 1.2281 & 2.3403 \\ 
\hline
\multirow{2}{*}{\textbf{marginal}} & \multicolumn{2}{c}{\textbf{500,000}} & & \multicolumn{2}{c}{\textbf{5,000,000}} & & \multicolumn{2}{c}{\textbf{500,000}} & & \multicolumn{2}{c}{\textbf{5,000,000}} \\ 
\cline{2-3} \cline{5-6} \cline{8-9} \cline{11-12} 
& $\sigma$ & IQR & & $\sigma$ & IQR & & $\sigma$ & IQR & & $\sigma$ & IQR \\ \hline
  10   & 0.0014 & 0.0018 &  & 0.0004 & 0.0006 &  & 0.0041 & 0.0020 &  & 0.0249 & 0.0270 \\ 
  20   & 0.0013 & 0.0018 &  & 0.0004 & 0.0006 &  & 0.0028 & 0.0020 &  & 0.0243 & 0.0280 \\ 
  100   & 0.0013 & 0.0017 &  & 0.0004 & 0.0005 &  & 0.0048 & 0.0040 &  & 0.0278 & 0.0442 \\ 
  500   & 0.0013 & 0.0017 &  & 0.0004 & 0.0005 &  & 0.0073 & 0.0040 &  & 0.0342 & 0.0540 \\ 
  1,000  & 0.0013 & 0.0017 &  & 0.0004 & 0.0005 &  & 0.0094 & 0.0050 &  & 0.0373 & 0.0560 \\ 
\hline
\multirow{2}{*}{\textbf{trapezium}}  & \multicolumn{2}{c}{\textbf{500,000}} & & \multicolumn{2}{c}{\textbf{5,000,000}} & & \multicolumn{2}{c}{\textbf{500,000}} & & \multicolumn{2}{c}{\textbf{5,000,000}} \\ 
\cline{2-3} \cline{5-6} \cline{8-9} \cline{11-12} 
 & $\sigma$ & IQR & & $\sigma$ & IQR & & $\sigma$ & IQR & & $\sigma$ & IQR \\ \hline
& 0.0013 & 0.0017 &  & 0.0004 & 0.0005 &  & 0.0081 & 0.0050 &  & 0.0371 & 0.0500 \\ 
\hline\hline
\end{tabular}
\end{table}

\textbf{Discussion of results} 
From Table \ref{tab: mean-med, mu=0.1, conc=50}, some general conclusions that are also valid in the other simulation settings, can be drawn about the bias. First of all, the bias of all approximations is low, even almost negligible. Overall, the bias of the $k$-means approximation and the trapezium approximation is smaller than the bias of the marginal approximation. The latter results in a bias that is not affected by the sample size of the data, while the bias of the $k$-means approximation depends on the sample size. As expected, the bias decreases when the number of clusters or boundaries increases. In general, we see that the bias based on the mean and the bias based on the median are close to each other. Hence, we can conclude that there were no extreme estimates. 

Table \ref{tab: mean-med, mu=0.1, conc=50} also reveals some general conclusions about the run time. Primarily, the marginal approximation results in the smallest run times, opposite to the $k$-means approximation with the largest run times. Hence, the run time of the trapezium approximation is always smaller than the one of the $k$-means approximation, but larger than the run time of the marginal approximation. As expected, the run time generally increases as the number of clusters or boundary values increases. The run time also increases as the sample size increases. More specifically, we see that when the sample size increases from $n=500,000$ to $n=5,000,000$, the run time gets approximately ten times larger for the marginal approximation and the trapezium approximation. However, for the $k$-means approximation, the increment factor of the run time is shorter than ten and becomes even shorter when the number of clusters increases. We also did one experiment with $n=50,000,000$, of which the results can be found in Table \ref{tab: sd-IQR, mu=0.95, conc=50, N=50,000,000} of Appendix B. There we see again that the run time increases when the number of clusters or boundary values increases. Comparing Table \ref{tab: sd-IQR, mu=0.95, conc=50, N=50,000,000} in Appendix B with Table \ref{tab: mean-med, mu=0.1, conc=50}, we also see a confirmation in the previous conclusion that the run time for the approximations increases when the sample size increases. Moreover, it is worth mentioning that even for this massive sample size, the run times remain brief. Finally, we see that the mean and median run time are in general nearly identical. Hence, we can conclude that there were no extreme run times.

The estimates vary little as can be seen from Table \ref{tab: sd-IQR, mu=0.1, conc=50}. From the simulations, we learn that the variability of the estimated concordance probability is higher for the $k$-means approximation than for the marginal approximation, at least for a low number of clusters or boundary values. This can be explained by the random starting values for the clusters in the $k$-means clustering algorithm, combined with the low number of clusters. For more than twenty clusters or boundary values, the standard deviation and the interquartile range are the same for the two approximations. Furthermore, the standard deviation and interquartile range do not change in function of the number of boundary values for the marginal approximation, whereas we do see a reduction of variability when the number of clusters increases for the $k$-means approximation. Overall, the variation of the estimates decreases for each of the four algorithms when the sample size increases. For the marginal approximation we can conclude that in general the run time variability increases when the sample size increases. This conclusion cannot be made for the $k$-means approximation.

Until here, only general conclusions of each specific simulation setting on its own are considered. To check whether there is an effect of the size of the population value on the algorithms, Tables \ref{tab: mean-med, mu=0.1, conc=50}, \ref{tab: mean-med, mu=0.1, conc=15} and \ref{tab: mean-med, mu=0.1, conc=150} are compared, as well as Tables \ref{tab: sd-IQR, mu=0.1, conc=50}, \ref{tab: sd-IQR, mu=0.1, conc=15} and \ref{tab: sd-IQR, mu=0.1, conc=150}. From these tables, we can see that the bigger the concentration (and hence the smaller the population value of the concordance probability), the smaller the bias of all approximations in most of the cases. Moreover, the true population value of the concordance probability has no effect on their run time and its variability for all algorithms. The true population value of the concordance probability has also no effect on the variability of their estimates. Another question was whether the extremity of the probability affects the algorithms. To answer this question, we compare the Tables \ref{tab: mean-med, mu=0.1, conc=50}, \ref{tab: mean-med, mu=0.5, conc=50} and  \ref{tab: mean-med, mu=0.25, conc=50}, as well as the Tables \ref{tab: sd-IQR, mu=0.1, conc=50}, \ref{tab: sd-IQR, mu=0.25, conc=50} and \ref{tab: sd-IQR, mu=0.5, conc=50}. The extremer the probability, the bigger the bias and the smaller run times of the marginal approximation. These conclusions cannot be drawn for the $k$-means approximation and the approximation based on the trapezium rule. For all algorithms, the extremity of the probability has no effect on the variability of the estimates and their run times. Finally, we can conclude that indeed only the relative size of the two groups matters, due to the small differences when comparing Table \ref{tab: mean-med, mu=0.05, conc=50} with Table \ref{tab: mean-med, mu=0.95, conc=50}, Table \ref{tab: mean-med, mu=0.25, conc=50} with Table \ref{tab: mean-med, mu=0.75, conc=50}, Table \ref{tab: sd-IQR, mu=0.05, conc=50} with Table \ref{tab: sd-IQR, mu=0.95, conc=50} and finally Table \ref{tab: sd-IQR, mu=0.25, conc=50} with Table \ref{tab: sd-IQR, mu=0.75, conc=50}. 

As a conclusion, this simulation study favors the marginal approximation over the approximations based on the $k$-means algorithm or the trapezium rule. The marginal approximation with 500 clusters or more delivers namely estimates that are very weakly biased. Moreover, this approximation is computed faster when dealing with the really large data sets, hence better passing the sample size complexity test than the $k$-means approximation or the one based on the trapezium rule. 
\FloatBarrier
\subsection{Continuous setting}
\label{sec: cont sim}
In this subsection, the goal is to investigate the performance of the $k$-means approximation and the marginal approximation, when the response variable is continuous. 

\textbf{Data generation setup}
We suppose that the observations and predictions both follow a standard normal distribution. In this simulation, the correlation between the observations and predictions varies between the following values $\rho \in \{0,0.25,0.5,0.75\}$. The parameter $\nu$ is determined such that $x$\% of the pairwise absolute differences of the observed values is smaller than $\nu$, with $x \in \{0, 20, 40\}$. 

A first step in this simulation is to determine the correct values for $\nu$. As these values are only based on the observations, they will be independent of the correlation $\rho$. Moreover, it is computationally impossible to precisely calculate these values for each sample and hence, the same values of $\nu$ will be consistently used. In order to determine these values, we first sample 10,000 observations from a standard normal distribution of which all pairwise absolute differences are calculated. The 20\% and 40\% percentiles of these absolute differences represent the values for $\nu$ in this specific sample, denoted by the pair $\mathbf{p}$. The above procedure is repeated 100 times, such that the means $\mathbf{q}$ of all obtained percentiles $\mathbf{p}_1,\ldots \mathbf{p}_{100}$ are a better estimate for the correct values of $\nu$. Since the ranges of $\mathbf{p}_1,\ldots \mathbf{p}_{100}$ can still be 3\%, the whole aforementioned procedure is once again repeated 100 times. The mean values of these obtained means $\mathbf{q}_1,\ldots \mathbf{q}_{100}$  are the final values for $\nu$, i.e. $\nu \in \{0, 0.3583, 0.7416\}$ for $x \in \{0, 20, 40\}$ respectively. These estimates are more reliable since the range of these means $\mathbf{q}_1,\ldots \mathbf{q}_{100}$ is maximum 0.4\%.

The next step for the simulation is to determine the population values of $C(\nu)$, for $\nu \in \{0, 0.3583, 0.7416\}$ and $\rho \in \{0,0.25,0.5,0.75\}$. For each combination of $\rho$ and $\nu$, the concordance probability is calculated by \eqref{ch1_cont_eq2} based on a sample of size 50,000. This is repeated 100 times, and the mean of these 100 concordance probabilities is denoted by $m$. Since the range of these concordance probabilities can still be 1.6\%, the above procedure is once again repeated five times. The mean value of the aforementioned obtained means $m_1, \ldots, m_5$ is the final value for $C(\nu)$. This estimation method is more reliable since the range of these means $m_1, \ldots, m_5$ is maximum 0.8\%. The final values of the concordance probability for each combination of $\rho$ and $\nu$ are denoted in Table \ref{tab: means}. Completely in line with our expectations, the concordance probability increases when the correlation between the predictions and observations increases. In case that $\rho$ is strictly positive, it can be seen that the concordance probability increases in function of $\nu$.

\begin{table}[hbt!]
\centering
\scriptsize
\caption{The population values of the concordance probability for several combinations of $\nu$ and $\rho$.}
\label{tab: means}
\begin{tabular}{ccccc}
\hline\hline
 && \multicolumn{3}{c}{$\nub$ }\\
 \cline{3-5}
$\rhob$ && 0 & 0.3583 & 0.7416 \\ 
  \hline
0 && 0.5000 &  0.5001   & 0.5001  \\ 
  0.25 && 0.5804 & 0.5973 & 0.6164  \\ 
  0.50 && 0.6666 & 0.7011 & 0.7387  \\ 
  0.75 && 0.7699 & 0.8233  & 0.8747  \\ 
   \hline \hline
\end{tabular}
\end{table}

\textbf{Evaluation setup}
Similar to the discrete setting, 1,000 samples are generated for each of the above described simulation settings. On these samples, the marginal approximation and the $k$-means approximation are applied to calculate the concordance probability. To our knowledge, we have proposed the first approximations to compute the concordance probability in the continuous setting and hence no benchmark methods are available. The $k$-means approximation tests the effect of 10, 20, 100, 500 or 1,000 clusters, whereas the marginal approximation focuses on 10, 20 or 100 boundary values. The latter only considers small numbers of boundary values due to the lengthy run times, as will be clear from the simulation results. Moreover, the boundary values are evenly spaced percentiles of the empirical distribution of the observed values, as advised in Section \ref{subsec: approx cont}. Focusing on the simulation setting with $\rho=0.25$ and $\nu = 0.3583$, Table \ref{tab: bias-time, rho=0.25 and nu=0.3583} shows the bias (based on the mean or median) together with the mean and median run time, whereas Table \ref{tab: sd-IQR, rho=0.25 and nu=0.3583} contains the standard deviation and the interquartile range of the computed concordance measure and run time. For the other simulation settings, the same tables are constructed as can be seen in Appendix C.
\begin{table}[ht]
\centering
\scriptsize
\caption{This table considers the continuous simulation setting with $\rho=0.25$ and $\nu=0.3583$. The mean and median bias and run time are shown in function of the number clusters for the $k$-means approximation and the number of boundary values for the marginal approximation. Two data set sizes are considered, namely 500,000 and 5,000,000.}
\label{tab: bias-time, rho=0.25 and nu=0.3583}
\begin{tabular}{cccccccccccc}
\hline\hline
 & \multicolumn{5}{c}{bias} & & \multicolumn{5}{c}{run time (s)}\\ 
 \cline{2-6} \cline{8-12} 
\multirow{2}{*}{\textbf{\textit{k}-means}}  & \multicolumn{2}{c}{\textbf{500,000}} & & \multicolumn{2}{c}{\textbf{5,000,000}} & & \multicolumn{2}{c}{\textbf{500,000}} & & \multicolumn{2}{c}{\textbf{5,000,000}} \\ 
\cline{2-3} \cline{5-6} \cline{8-9} \cline{11-12} 
&  mean & median & & mean & median & & mean & median & & mean & median \\ \hline
10 &   0.0181 & 0.0173 &  & 0.0208 & 0.0196 &  & 1.0409 & 0.9020 &  & 6.8401 & 5.4290 \\ 
  20 & 0.0077 & 0.0076 &  & 0.0097 & 0.0096 &  & 1.1936 & 0.9900 &  & 7.3614 & 5.9335 \\ 
  100   & 0.0014 & 0.0015 &  & 0.0016 & 0.0016 &  & 2.2619 & 1.6825 &  & 11.4168 & 10.3470 \\ 
  500   & 0.0003 & 0.0003 &  & 0.0003 & 0.0003 &  & 14.3973 & 14.0835 &  & 26.5740 & 26.1695 \\ 
  1,000  & 0.0002 & 0.0002 &  & 0.0002 & 0.0002 &  & 33.1026 & 34.3245 &  & 45.5225 & 45.2805 \\ 
\hline
\multirow{2}{*}{\textbf{marginal}} & \multicolumn{2}{c}{\textbf{500,000}} & & \multicolumn{2}{c}{\textbf{5,000,000}} & & \multicolumn{2}{c}{\textbf{500,000}} & & \multicolumn{2}{c}{\textbf{5,000,000}} \\ 
\cline{2-3} \cline{5-6} \cline{8-9} \cline{11-12} 
&  mean & median & & mean & median & & mean & median & & mean & median \\ \hline
  10 & 0.0260 & 0.0260 &  & 0.0259 & 0.0259 &  & 2.0783 & 2.0770 &  & 26.8491 & 26.8590 \\ 
  20 & 0.0119 & 0.0119 &  & 0.0119 & 0.0119 &  & 3.8648 & 3.8660 &  & 49.2721 & 49.2805 \\ 
  100   & 0.0028 & 0.0028 &  & 0.0028 & 0.0028 &  & 85.7858 & 85.7620 &  & 316.6178 & 315.9965 \\ 
\hline\hline
\end{tabular}
\end{table}

\textbf{Discussion of results} 
From Table \ref{tab: bias-time, rho=0.25 and nu=0.3583}, some general conclusions that are also valid in the other continuous simulation settings, can be made about the bias. The bias is low for all approximations and there is a decrease in the bias when the number of clusters or boundaries increases. Moreover, similar to the discrete setting, the bias of the $k$-means approximation is smaller than the one of the marginal approximation when using the same number of clusters or boundaries. The sample size of the data has clearly no effect on the bias of the marginal approximation, as well as on the $k$-means approximation with 100 clusters or more. 

Finally, when the number of clusters is larger than ten, we see that the bias based on the mean and the bias based on the median are almost similar. Hence, we can conclude that no extreme estimates were encountered in these cases. The same holds for all considered numbers of boundaries for the marginal approximation. A note of caution is due here, since some aforementioned conclusions are not valid in case there is no correlation between the observations and the predictions (see Appendix C). Firstly, the bias of the marginal approximation seems independent of the number of boundaries. Secondly, the bias of the $k$-means approximation is only smaller than the one of the marginal approximation in case of less than 100 clusters or boundary values. Once it is higher than 100, the bias of both approximations is equal.

Table \ref{tab: bias-time, rho=0.25 and nu=0.3583} also reveals some general conclusions about the run time. In contrast to the discrete setting, the run time of the marginal approximation is higher than the one of the $k$-means approximation. More specifically, the $k$-means method is approximately two times faster to compute than the marginal method for ten clusters or boundary values, while for hundred clusters or boundary values, it is thirty to fifty times faster. Next, we see that the run time generally increases as the number of clusters or boundary values increases. Note that sometimes, this is not the case when we compare the simulation results for 10 and 20 clusters. Furthermore, the run time increases as the sample size increases for both approximations. Finally, there are no extreme run times, since the mean and median run times are nearly identical. 

\begin{table}[ht]
\centering
\scriptsize
\caption{This table considers the continuous simulation setting with $\rho=0.25$ and $\nu=0.3583$. The interquartile range and standard deviation over the estimates and over the run time are shown in function of the number of boundary values for the marginal approximation and the number of clusters for the $k$-means approximation. Two data set sizes are considered, namely 500,000 and 5,000,000.}
\label{tab: sd-IQR, rho=0.25 and nu=0.3583}
\begin{tabular}{cccccccccccc}
\hline\hline
 & \multicolumn{5}{c}{$\hat{C}$} & & \multicolumn{5}{c}{run time (s)}\\ 
 \cline{2-6} \cline{8-12} 
\multirow{2}{*}{\textbf{\textit{k}-means}}  & \multicolumn{2}{c}{\textbf{500,000}} & & \multicolumn{2}{c}{\textbf{5,000,000}} & & \multicolumn{2}{c}{\textbf{500,000}} & & \multicolumn{2}{c}{\textbf{5,000,000}} \\ 
\cline{2-3} \cline{5-6} \cline{8-9} \cline{11-12} 
& $\sigma$ & IQR & & $\sigma$ & IQR & & $\sigma$ & IQR & & $\sigma$ & IQR \\ \hline
10 &  0.0342 & 0.0447 &  & 0.0338 & 0.0436 &  & 0.5161 & 0.7095 &  & 2.3342 & 3.6027 \\ 
  20 & 0.0165 & 0.0235 &  & 0.0160 & 0.0220 &  & 0.6839 & 0.8413 &  & 2.4344 & 3.9245 \\ 
  100 & 0.0026 & 0.0033 &  & 0.0022 & 0.0029 &  & 1.4296 & 1.8160 &  & 2.6989 & 0.3110 \\ 
  500 & 0.0006 & 0.0009 &  & 0.0003 & 0.0004 &  & 9.3571 & 18.3248 &  & 2.5966 & 0.2913 \\ 
  1,000 & 0.0005 & 0.0007 &  & 0.0002 & 0.0003 &  & 16.8481 & 29.5448 &  & 2.8324 & 0.3230 \\ 
\hline
\multirow{2}{*}{\textbf{marginal}} & \multicolumn{2}{c}{\textbf{500,000}} & & \multicolumn{2}{c}{\textbf{5,000,000}} & & \multicolumn{2}{c}{\textbf{500,000}} & & \multicolumn{2}{c}{\textbf{5,000,000}} \\ 
\cline{2-3} \cline{5-6} \cline{8-9} \cline{11-12} 
& $\sigma$ & IQR & & $\sigma$ & IQR & & $\sigma$ & IQR & & $\sigma$ & IQR \\ \hline
  10 & 0.0007 & 0.0009 &  & 0.0002 & 0.0003 &  & 0.0516 & 0.0790 &  & 1.8140 & 2.5202 \\ 
  20 & 0.0006 & 0.0008 &  & 0.0002 & 0.0003 &  & 0.0632 & 0.1060 &  & 2.9274 & 3.7165 \\ 
  100  & 0.0005 & 0.0007 &  & 0.0002 & 0.0002 &  & 0.7199 & 1.0315 &  & 8.9207 & 11.7882 \\ 
\hline\hline
\end{tabular}
\end{table}

The estimates vary little as can be seen in Table \ref{tab: sd-IQR, rho=0.25 and nu=0.3583}. From the simulations, we learn that the variability of the estimated concordance probability is higher for the $k$-means approximation than for the marginal approximation when the same number of clusters and boundary values is considered. The $k$-means approximation with more than 100 clusters has approximately the same values for the standard deviation and interquartile range of the estimates, as the marginal approximation with 10, 20 or 100 boundary values. This was expected due to the random starting values for the clusters in the $k$-means clustering algorithm. The latter also explains why we see a reduction of the variability when the number of clusters increases for the $k$-means approximation. However, for the marginal approximation the standard deviation and interquartile range of the estimates hardly change in function of the number of boundary values. Overall, the variation of the estimates decreases when the sample size increases. 

Thus far, the focus was on the variability of the estimates. In Table \ref{tab: sd-IQR, rho=0.25 and nu=0.3583}, the variability of the run times is also studied. When the number of observations equals 500,000, the variability of the run time is mostly higher for the $k$-means approximation than for the marginal approximation. Moreover, it increases when the number of clusters or boundary values increases. For 5,000,000 observations, we still see an increase of variability when the number of clusters or boundary values increases, apart from the interquartile range for the cases with 10 or 20 clusters. Moreover, the variability of the run time is in this case typically smaller for the $k$-means approximation than for the marginal approximation. Going from 500,000 to 5,000,000 observations leads to an increase in the variability of the run time for the marginal approximation. For the $k$-means approximation, the same holds for the standard deviation (interquartile range) in case of 100 (20) clusters or less. However, a larger number of clusters results in a decrease of variability in the run time when going from 500,000 observations to 5,000,000 observations.

Finally, we also consider the effect of $\rho$ and $\nu$ on both approximations. In general, the bias increases when the correlation increases, but it is independent of $\nu$. Focusing on the run time, we see that it seems to be unaffected by the correlation between the observed and predicted values. As expected, the run time decreases for increasing values of $\nu$, since larger values of $\nu$ result in smaller numbers of pairs that should be compared. Next, the variability of the run times and the estimates does not change in function of $\nu$. However, the variability of the estimates decreases when $\rho$ increases, whereas there is no link between the variability of the run times and $\rho$.

To conclude, this simulation study clearly favors the $k$-means approximation over the marginal one. The $k$-means approximation delivers namely estimates that are very weakly biased, even if the number of clusters is poorly tuned. Also, the $k$-means approximation is computed much faster when dealing with really large data sets, hence better passing the sample size complexity test than the marginal approximation. As a general recommendation, one should choose 100 clusters when using the $k$-means approximation, as the fitting time is still very small, and as this results in a low variability over the estimated concordance probability.
\section{Real-life examples}
\label{sec: ex}
In this section, two real life data sets from Kaggle\footnote{https://www.kaggle.com} are used to compare the exact concordance probability with the approximations discussed in this paper. The first example focuses on the discrete setting, whereas the second one deals with a continuous response variable. Both examples are executed on a computer with specifications Intel Core i7-8650U CPU
@ 1.90GHz 2.11GHz processor.
\FloatBarrier
\subsection{Predict Click through rate (CTR) for a website}
We consider data about the click through rate (CTR) for a website related to job searches \citep{CTR}. This data set consists of 10 variables and 1,200,890 observations, where each observation corresponds to a user's view of a job listing. One of the variables is called \verb|apply|, which indicates whether or not the user has applied for this job listing after checking the job description.  The related notebook ``Let's Start'' on Kaggle cleans the data first, then splits the data set into a training and test set (respectively containing 871,290 and 81,704 observations) and finally applies three popular predictive models: XGBoost (xgb), random forest (rf) and a Light Gradient Boosted Machine (lgbm). The main interest of this paper is on the calculation of the concordance probability of the model once the predictions are available. Normally, this measure is only determined for the test set, but since we also focus on large data sets, we consider the predictions of the entire cleaned data set as well.  

In order to determine the bias of the concordance probability estimates, we first need to determine its exact value. Therefore, the data set is first split into a zero-group and a one-group, based on the value of the variable \verb|apply|. In this case the zero-group is the smallest, which is why a for loop is used for the elements in this group. In each iteration, we count the number of predictions in the one-group that are larger than the prediction of the considered element of the zero-group. Summing up all these counts, divided by the number of considered pairs, results in the exact concordance probability. The latter can be found in Table \ref{tab: ex CTR exact} and shows that, based on the concordance probability, the third model based on a Light Gradient Boosted Machine, performs best on the test set. Moreover, it requires an excessive amount of time to calculate the concordance probability on the complete cleaned data set.
\begin{table}[hbt!]
\centering
\scriptsize
\caption{The first three columns show the exact concordance probabilities obtained for the three different models of the CTR example, for the test set (test) or the complete cleaned data set (all). In the last three columns, the computing times are shown.}
\label{tab: ex CTR exact}
\begin{tabular}{ccccccccc}
  \hline\hline
 && \multicolumn{3}{c}{$C$} && \multicolumn{3}{c}{run time (s)} \\ 
 \cline{3-5} \cline{7-9} 
 && lgbm & xgb & rf && lgbm & xgb & rf \\
  \hline
test && 0.5842 & 0.5838 & 0.5655 && 3.58 & 3.64 & 3.66 \\ 
  all && 0.5940 & 0.5872 & 0.7052 && 428.20 & 432.33 & 427.16 \\ 
   \hline\hline
\end{tabular}
\end{table}

Table \ref{tab: ex CTR approx} shows for the three predictive models the bias of the different approximations for the concordance probability discussed in this paper. First of all, the bias of all approximations is extremely low, even almost negligible. More specifically, the smallest bias is in this example obtained by approximations based not only on the trapezium rule, but also on the $k$-means algorithm with 1,000 clusters. However, the approximations based on the $k$-means algorithm do have the largest run time. The smallest run times are obtained by the marginal approximation. Moreover, comparing Tables \ref{tab: ex CTR exact} and \ref{tab: ex CTR approx} show that all approximations for the complete cleaned data set have a much smaller run time than the one of the exact calculation. Finally, general conclusions from the simulation study in the previous section are confirmed in this example, e.g. the bias decreases when the number of boundaries increases, the run time increases when the number of clusters increases, the run time increases when the number of observations increases and so on.

\begin{sidewaystable}[hbt!]
\centering
\scriptsize
\caption{In the left part of this table, we can see the bias of the concordance probability estimates for the CTR example, based on the $k$-means approximation, the marginal approximation and the one based on the trapezium rule. The right part shows the corresponding run times of these approximations.}
\label{tab: ex CTR approx}
\begin{tabular}{cccccccccccccccc}
\hline\hline
& \multicolumn{7}{c}{bias} && \multicolumn{7}{c}{run time (s)}\\
 \cline{2-8} \cline{10-16} 
 \multirow{2}{*}{\textbf{\textit{k}-means}} & \multicolumn{3}{c}{test} & & \multicolumn{3}{c}{all}& & \multicolumn{3}{c}{test} & & \multicolumn{3}{c}{all}\\ 
 \cline{2-4} \cline{6-8} \cline{10-12} \cline{14-16} 
  &  lgbm & xgb & rf && lgbm & xgb & rf&& lgbm & xgb & rf&& lgbm & xgb & rf\\ 
\hline
10 &   -0.0193 & -0.0365 & 0.0359 &  & 0.0096 & -0.0273 & 0.0069 &  & 0.12 & 0.17 & 0.10 &  & 1.61 & 1.89 & 1.08 \\ 
  20   & -0.0074 & 0.0067 & -0.0129 &  & 0.0036 & 0.0025 & 0.0017 &  & 0.31 & 0.26 & 0.22 &  & 1.55 & 2.23 & 2.89 \\ 
  100   & 0.0004 & -0.0004 & -0.0001 &  & 0.0003 & 0.0001 & -0.0002 &  & 0.48 & 0.48 & 0.27 &  & 2.58 & 4.31 & 1.67 \\ 
  500   & 0.0001 & 0.0001 & 0.0000 &  & 0.0000 & -0.0000 & -0.0001 &  & 4.65 & 4.77 & 5.36 &  & 14.16 & 13.75 & 9.17 \\ 
  1,000   & 0.0000 & 0.0001 & 0.0000 &  & 0.0001 & 0.0000 & 0.0000 &  & 18.46 & 17.61 & 18.17 &  & 38.42 & 38.24 & 25.06 \\
\hline
 \multirow{2}{*}{\textbf{marginal}} & \multicolumn{3}{c}{test} & & \multicolumn{3}{c}{all}& & \multicolumn{3}{c}{test} & & \multicolumn{3}{c}{all}\\ 
 \cline{2-4} \cline{6-8} \cline{10-12} \cline{14-16} 
  &  lgbm & xgb & rf && lgbm & xgb & rf&& lgbm & xgb & rf&& lgbm & xgb & rf\\ 
\hline
10 &  0.0077 & 0.0074 & 0.0063 &  & 0.0084 & 0.0077 & 0.0165 &  & 0.03 & 0.03 & 0.03 &  & 0.19 & 0.19 & 0.22 \\ 
  20 & 0.0040 & 0.0041 & 0.0031 &  & 0.0043 & 0.0041 & 0.0089 &  & 0.05 & 0.03 & 0.03 &  & 0.22 & 0.22 & 0.21 \\ 
  100  & 0.0010 & 0.0009 & 0.0007 &  & 0.0010 & 0.0009 & 0.0019 &  & 0.03 & 0.03 & 0.03 &  & 0.21 & 0.18 & 0.23 \\ 
  500 & 0.0003 & 0.0003 & 0.0002 &  & 0.0003 & 0.0002 & 0.0004 &  & 0.04 & 0.06 & 0.06 &  & 0.40 & 0.25 & 0.39 \\ 
  1,000  & 0.0002 & 0.0002 & 0.0001 &  & 0.0002 & 0.0001 & 0.0002 &  & 0.08 & 0.08 & 0.11 &  & 0.42 & 0.37 & 0.50 \\  
\hline
 \multirow{2}{*}{\textbf{trapezium}} & \multicolumn{3}{c}{test} & & \multicolumn{3}{c}{all}& & \multicolumn{3}{c}{test} & & \multicolumn{3}{c}{all}\\ 
  \cline{2-4} \cline{6-8} \cline{10-12} \cline{14-16} 
 &  lgbm & xgb & rf && lgbm & xgb & rf&& lgbm & xgb & rf&& lgbm & xgb & rf\\ 
\hline
    & 0.0001 & 0.0001 & 0.0000 &  & 0.0000 & 0.0000 & 0.0000 &  & 0.07 & 0.04 & 0.05 &  & 0.37 & 0.41 & 0.55  \\ 
\hline
\hline
\end{tabular}
\end{sidewaystable}
\FloatBarrier
\subsection{New York City taxi fare prediction}
We consider data about the taxi fares in New York \citep{taxi}. This data set consists of a train and test set of respectively 55,423,856 and 9,914 observations, representing taxi rides in New York City, and seven variables each. One of the variables is called \verb|fare_amount|, which is a continuous variable indicating the fare amount (inclusive of tolls) for a taxi ride in New York City. The other six variables represent the pickup and drop off locations, the pickup time and the number of passengers. The related notebook ``NYC Taxi Fare - Data Exploration'' written by \cite{taxiNotebook} uses 2,000,000 rows of the training data to construct a linear regression model in order to predict the variable \verb|fare_amount|. The first step is cleaning the data, e.g. removing missing data, removing noisy data points with underwater locations and so on. This results into a data set consisting of 1,918,905 lines, of which 75\% will be used to train the model and 25\% to test the model. In the next step, some data exploration reveals that the important features will be the year and hour of the pickup time, the driven distance and the number of passengers. The latter variables are therefore used in the linear regression model to predict the fare amount for a taxi ride in New York City. Once again, the interest of this paper is not on how this model is constructed, but on the calculation of the concordance probability of the model once the predictions are available. Normally, this measure is only determined for the test set which consists here of 479,727 observations. Since we focus on large data sets, we also consider the 1,918,905 predictions of the entire cleaned data set. Note that the correlation between the predictions and the observations is 89.52\% for the test set and 89.24\% for the complete cleaned data set.

The goal is to predict the C-index for different values of $\nu$. Similar to the simulation study, we therefore first determine the value of $\nu$ such that $x$\% of the pairwise absolute differences of the observed values is smaller than $\nu$, with $x \in \{ 0, 10, 20, 30, 40, 50\}$. A first possible way to determine these values is by considering all possible pairs between the observations in order to determine the absolute differences between observations belonging to the same pair. However, in case of 1,918,905 observations, this would result in 1,841,097,240,060 pairs and corresponding differences. A closer look to the data reveals that out of these almost two million observations, only 2,569 unique values for the fare amount for a taxi ride in NYC are denoted. These unique values are represented in $\mathbf{y}$ and the corresponding frequencies in $\mathbf{w}$. Applying the function \verb|combn| in \verb|R| on $\mathbf{y}$, results in all possible pairs of these unique values. The corresponding frequencies of these pairs are obtained by the strict upper triangle of the matrix $\mathbf{w}\mathbf{w}^T$. However, do note that there are still $\frac{1}{2}\mathbf{w}^T(\mathbf{w}-1)$ pairs possible that consist of two identical observations, and hence with a difference of zero between both. After summing up the frequencies of pairs with the same absolute difference between their observations, the empirical cumulative distribution function of those differences can be constructed. From this function, the $\nu$ values are determined and represented in Table \ref{tab: taxi nu}.

\begin{table}[hbt!]
\centering
\scriptsize
\caption{The values for $\nu$ such that $x$\% of the absolute differences between the observed values is smaller than $\nu$. This is done for the entire data set (all) and the test set (test).}
\label{tab: taxi nu}
\begin{tabular}{ccccccccc}
  \hline\hline
  &&\multicolumn{6}{c}{$x$}\\
  \cline{3-9}
 && 0\% & 10\% & 20\% & 30\% & 40\% & 50\% \\
  \hline
test && 0 & 12.12 & 20.75 & 25.60 & 31.64 & 38.54\\
all && 0 & 12.87 & 22.20 & 30.30 & 39.44 & 49.43 \\ 
   \hline\hline
\end{tabular}
\end{table}

In order to determine the bias of the concordance probability estimates, we first need to determine its exact value. Note that we cannot take advantage anymore of the small number of unique values in the observations, since their predictions can differ. Therefore, a for loop is used for all the observations. In each iteration, we select the rows with an observation strictly larger than the considered observation added up with $\nu$. The number of selected rows is for each iteration stored in $\mathbf{u}$.  In this selection, we count the  number of predictions that are larger than the prediction of the considered element, and store this value in $\mathbf{v}$. Hence, the concordance probability can be obtained by the division of $\bar{\mathbf{v}}$ by $\bar{\mathbf{u}}$. For all considered values of $\nu$, the concordance probability for the test set and the complete data set can be found in Table \ref{tab: ex taxi exact}. Note that the run times to calculate the exact concordance probability on the complete cleaned data set are staggering.

\begin{table}[hbt!]
\centering
\scriptsize
\caption{ This table shows the exact concordance probabilities together with the computing times for different values for $\nu$. The upper part focuses on the test set (test) and the lower part on the complete cleaned data set (all).}
\label{tab: ex taxi exact}
\begin{tabular}{cccccccc}
  \hline\hline
 && \multicolumn{6}{c}{$\nu$ - test }  \\ 
 \cline{3-8} 
 && 0 & 12.12 & 20.75 & 25.60 & 31.64 & 38.54  \\
  \hline
$C$ && 0.8583 & 0.9808 & 0.9817 & 0.9800 & 0.9773 & 0.9783  \\ 
  run time (s) && 8,908 & 4,125 & 3,455 & 3,295 & 3,159 & 3,023 \\ 
  &&&&&&&\\
  && \multicolumn{6}{c}{$\nu$ - all}\\ 
 \cline{3-8} 
   && 0 & 12.87 & 22.20 & 30.30 & 39.44 & 49.43\\
  \hline
 $C$ && 0.8577 & 0.9801 & 0.9799 & 0.9763 & 0.9772 & 0.9810\\
  run time (s) &&  138,058 & 58,355  & 45,559   & 36,842 & 31,896 & 27,397\\
   \hline\hline
\end{tabular}
\end{table}

\begin{sidewaystable}[hbt!]
\centering
\scriptsize
\caption{The bias of the concordance probability estimates for the taxi fare example, based on the $k$-means approximation and the marginal approximation. This is done for the test set (test) and the complete data set (all), for different values of $\nu$.}
\label{tab: ex taxi approx bias}
\begin{tabular}{cccccccccccccc}
\hline\hline
 & \multicolumn{13}{c}{bias} \\ 
 \cline{2-14}
\multirow{2}{*}{\textbf{\textit{k}-means}} & \multicolumn{6}{c}{$\nu$ - test}  & & \multicolumn{6}{c}{$\nu$ - all} \\
\cline{2-7} \cline{9-14}
& 0 & 12.12 & 20.75 & 25.60 & 31.64 & 38.54 && 0 & 12.87 & 22.20 & 30.30 & 39.44 & 49.43\\
\hline
10 & 0.1405 & 0.0192 & 0.0183 & 0.0200 & 0.0227 & 0.0212 &  & 0.1413 & 0.0199 & 0.0201 & 0.0237 & 0.0228 &  $-$  \\ 
  20 & 0.0398 & 0.0147 & 0.0113 & 0.0200 & 0.0056 & 0.0163 &  & 0.0386 & 0.0196 & 0.0169 & 0.0237 & 0.0228 & 0.0190 \\ 
  100 & 0.0028 & 0.0099 & 0.0094 & 0.0100 & 0.0094 & 0.0142 &  & 0.0020 & 0.0055 & 0.0087 & 0.0067 & 0.0079 & 0.0190 \\ 
  500 & -0.0069 & 0.0036 & 0.0042 & 0.0028 & 0.0068 & 0.0101 &  & -0.0071 & 0.0026 & 0.0065 & 0.0082 & 0.0068 & 0.0188 \\ 
  1,000 & -0.0076 & 0.0019 & 0.0015 & 0.0045 & 0.0029 & 0.0053 &  & -0.0083 & 0.0021 & 0.0033 & 0.0028 & 0.0099 & 0.0149 \\ 
\multirow{2}{*}{\textbf{marginal}} & \multicolumn{6}{c}{$\nu$ - test}  & & \multicolumn{6}{c}{$\nu$ - all} \\
\cline{2-7} \cline{9-14}
& 0 & 12.12 & 20.75 & 25.60 & 31.64 & 38.54 && 0 & 12.87 & 22.20 & 30.30 & 39.44 & 49.43\\
\hline
10& 0.0534 & 0.0052 & 0.0072 & 0.0147 & $-$  & $-$  &  & 0.0538 & 0.0035 & 0.0103& $-$ & $-$ & $-$ \\ 
  20& 0.0319 & 0.0055 & 0.0065 & 0.0141 &  & $-$  & $-$  & 0.0325 & 0.0051 & 0.0090  & $-$ & $-$ & $-$ \\ 
  100& 0.0123 & 0.0024 & 0.0010 & 0.0005 & 0.0012 & 0.0044 &  & 0.0135 & 0.0022 & 0.0010 & 0.0006 & 0.0045 & $-$ \\ 
  500& 0.0091 & 0.0011 & 0.0008 & 0.0007 & 0.0008 & 0.0012 &  & 0.0101 & 0.0010 & 0.0008 & 0.0007 & 0.0013 & 0.0042  \\ 
  1,000& 0.0075 & 0.0008 & 0.0006 & 0.0006 & 0.0007 & 0.0010 &  & 0.0087 & 0.0007 & 0.0006 & 0.0007 & 0.0011 & 0.0026 \\ 
  \hline\hline
\end{tabular}
\end{sidewaystable}

\begin{sidewaystable}[hbt!]
\centering
\scriptsize
\caption{The  run times corresponding to the calculation of the concordance probability estimates for the taxi fare example, based on the $k$-means approximation and the marginal approximation. This is done for the test set (test) and the complete data set (all), for different values of $\nu$.}
\label{tab: ex taxi approx time}
\begin{tabular}{cccccccccccccc}
\hline\hline
 & \multicolumn{13}{c}{run time (s)} \\ 
 \cline{2-14}
\multirow{2}{*}{\textbf{\textit{k}-means}} & \multicolumn{6}{c}{$\nu$ - test}  & & \multicolumn{6}{c}{$\nu$ - all} \\
\cline{2-7} \cline{9-14}
& 0 & 12.12 & 20.75 & 25.60 & 31.64 & 38.54 && 0 & 12.87 & 22.20 & 30.30 & 39.44 & 49.43\\
\hline
10& 1.10& 1.14& 1.89& 1.83& 1.72& 1.30&  & 5.14& 3.57& 6.19& 5.69& 2.24& 9.12\\ 
  20& 1.74& 3.15& 4.63& 2.52& 3.19& 2.15&  & 2.40& 3.01& 7.02& 6.75& 7.02& 11.31 \\ 
  100& 1.22& 0.97& 1.70& 2.01& 2.06& 0.69&  & 5.75& 5.06& 5.04& 3.55& 10.48& 3.45 \\ 
  500& 4.07& 2.47& 3.91& 3.38& 2.33& 2.51&  & 11.86& 8.46& 10.41& 9.18& 8.28& 7.99 \\ 
  1,000& 9.46& 5.31& 6.50& 4.20& 6.45& 4.79&  & 25.00& 14.58& 14.64& 14.29& 13.36& 13.81 \\ 
\multirow{2}{*}{\textbf{marginal}} & \multicolumn{6}{c}{$\nu$ - test}  & & \multicolumn{6}{c}{$\nu$ - all} \\
\cline{2-7} \cline{9-14}
& 0 & 12.12 & 20.75 & 25.60 & 31.64 & 38.54 && 0 & 12.87 & 22.20 & 30.30 & 39.44 & 49.43\\
\hline

  10& 2.66& 3.16& 3.17& 2.99& 3.03& 4.18&  & 11.81& 13.49& 14.18& 13.19& 13.63& 13.11 \\ 
  20& 4.58 & 6.95& 6.02& 6.20& 5.68& 6.09&  & 18.33& 25.71& 27.26& 24.25& 26.41& 24.90 \\ 
  100& 14.38 & 20.92& 20.17& 19.08& 19.06& 18.97&  & 58.09& 82.01& 79.48& 79.64& 81.29& 84.61 \\ 
  500& 29.00& 37.63& 38.85& 36.17& 36.06& 35.78&  & 115.20& 152.52& 152.51& 145.28& 142.08& 150.61 \\ 
  1,000& 38.86& 47.19& 43.99& 44.92& 43.51& 44.14&  & 150.41& 173.61& 174.67& 179.13& 179.27 & 179.15 \\ 
  \hline\hline
\end{tabular}
\end{sidewaystable}
Tables \ref{tab: ex taxi approx bias} and \ref{tab: ex taxi approx time} show respectively the bias and run time of the marginal approximation and the $k$-means approximation in this example. First of all, the bias of all approximations is low, especially for the marginal approximation when $\nu>0$. Nevertheless, we still advise to use the $k$-means approximation in this continuous version due to the very small run times, only yielding a small bias. Moreover, the marginal algorithm is not always able to determine an approximation when $\nu$ is large and the number of boundary value is small. In these cases, the low number of regions makes it namely impossible to compare regions that satisfy the restriction of $\nu$. Focusing for example on the complete data set with $100$ boundary values and $\nu=49.43$, the smallest boundary value $\tau_1$ is 3.3 whereas the biggest boundary value $\tau_{100}$ is 52. Since the absolute difference between $\tau_{100}$ and $\tau_1$ is smaller than $\nu$, no valid region-to-region comparisons can be made. Despite the fact that the run times of the marginal algorithm are much higher than the ones of the $k$-means algorithm, it is worth mentioning that they are still much smaller than the original ones denoted in Table \ref{tab: ex taxi exact}. Finally, general conclusions from the simulation study are confirmed in this example, e.g. the bias decreases when the number of boundaries increases, the run time increases when the sample size increases and so on.

\FloatBarrier

\section{Conclusion}
The current existing algorithms to compute the concordance probability are only adapted to a discrete response variable in case of large data sets. Therefore, we proposed two computationally efficient algorithms to estimate the concordance probability in both the discrete and continuous setting: the $k$-means algorithm and the marginal algorithm. In the discrete setting, the marginal approximation is the fastest method and moreover, it achieves the same bias as the standard approximation based on the trapezium rule. However, in case of a continuous response variable, the $k$-means algorithm is the most precise and results in the smallest run time. These conclusions are based on an extensive simulation study, that also focuses on the variability of the estimates and the run times of the algorithms. The good performance is also illustrated on two real data applications. Further lines of research can consist of adapting these methods to different types of concordance probabilities. The implementations of the discussed approximations of the concordance probability are written in \verb R  and will become available on github and as an \verb R -package.

\appendix
\section*{Appendix A.}
\label{app:betabin}
Assume that $X$ is a random variable following a beta-binomial distribution, e.g. $X\sim BB(\alpha,\beta,n)$, and denote the probability density function of the beta-binomial, beta and binomial distribution respectively as $f_{BB}, f_{Beta}$ and $f_{Bin}$. Then the following calculations show the relation between these distributions as discussed in Section~\ref{sec: disc sim}:
\begin{eqnarray}
f_{BB}(x|\alpha, \beta, n) & = & \int_0^1 f_{Bin}(x|p) f_{Beta}(p|\alpha,\beta)dp\nonumber\\
&=& \int_0^1 \binom{n}{x}p^x(1-p)^{n-x}\frac{p^{\alpha-1}(1-p)^{\beta-1}}{Beta(\alpha,\beta)}dp\nonumber\\
&=& \frac{\binom{n}{x}}{Beta(\alpha,\beta)}\int_0^1p^{x+\alpha-1}(1-p)^{n-x+\beta-1}dp\\
&=& \frac{\binom{n}{x}}{Beta(\alpha,\beta)} Beta(x+\alpha,n-k+\beta)\\
&=& \frac{\Gamma(n+1)\Gamma(\alpha+\beta)\Gamma(x+\alpha)\Gamma(n-x+\beta)}{\Gamma(x+1)\Gamma(n-x+1)\Gamma(\alpha)\Gamma(\beta)\Gamma(n+\alpha+\beta)}.
\end{eqnarray}
From Equation (1) to Equation (2), the definition of the beta function $Beta$ is used, and from Equation (2) to (3) the definition of the gamma function $\Gamma$ together with its relation to the beta function.\\

\section*{Appendix B.}
In this appendix, the results of the simulation study in the discrete setting are shown.

\begin{table}[ht]
\centering
\scriptsize
\caption{This table considers the discrete simulation setting with $\mu=0.50$ and $\kappa=50$. The mean and median bias and run time are shown for the approximation based on the trapezium rule, the marginal approximation (in function of the number of boundaries) and the $k$-means approximation (in function of the number of clusters). Two data set sizes are considered, namely 500,000 and 5,000,000.
}
\label{tab: mean-med, mu=0.5, conc=50}
\begin{tabular}{ccccccccccccc}
\hline\hline
 & \multicolumn{5}{c}{bias} & & \multicolumn{5}{c}{run time (s)}\\ 
 \cline{2-6} \cline{8-12} 
\multirow{2}{*}{\textbf{\textit{k}-means}}  & \multicolumn{2}{c}{\textbf{500,000}} & & \multicolumn{2}{c}{\textbf{5,000,000}} & & \multicolumn{2}{c}{\textbf{500,000}} & & \multicolumn{2}{c}{\textbf{5,000,000}} \\ 
\cline{2-3} \cline{5-6} \cline{8-9} \cline{11-12} 
& mean & median & & mean & median & & mean & median & & mean & median \\ \hline
10  & -0.0075 & -0.0107 &  & -0.0021 & -0.0023 &  & 0.5167 & 0.5070 &  & 4.6971 & 4.6490 \\ 
  20   & -0.0001 & 0.0002 &  & 0.0001 & 0.0002 &  & 0.6277 & 0.6065 &  & 4.6992 & 4.5465 \\ 
  100   & 0.0000 & 0.0000 &  &  0.0000 &  0.0000 &  & 0.8426 & 0.8810 &  & 6.3347 & 6.3110 \\ 
  500   & 0.0000 & 0.0000 &  & 0.0000 & 0.0000 &  & 4.9459 & 4.7380 &  & 19.7852 & 19.7825 \\ 
  1,000   & 0.0000 & 0.0000 &  & 0.0000 & 0.0000 &  & 16.5195 & 16.0410 &  & 40.8346 & 40.8205 \\
\hline\hline
\end{tabular}
\end{table}

\begin{table}[ht]
\centering
\scriptsize
\caption{This table considers the discrete simulation setting with $\mu=0.10$ and $\kappa=15$. The mean and median bias and run time are shown for the approximation based on the trapezium rule, the marginal approximation (in function of the number of boundaries) and the $k$-means approximation (in function of the number of clusters). Two data set sizes are considered, namely 500,000 and 5,000,000.}
\label{tab: mean-med, mu=0.1, conc=15}
\begin{tabular}{cccccccccccc}
\hline\hline
 &  \multicolumn{5}{c}{bias} & & \multicolumn{5}{c}{run time (s)}\\ 
 \cline{2-6} \cline{8-12} 
\multirow{2}{*}{\textbf{\textit{k}-means}}  & \multicolumn{2}{c}{\textbf{500,000}} & & \multicolumn{2}{c}{\textbf{5,000,000}} & & \multicolumn{2}{c}{\textbf{500,000}} & & \multicolumn{2}{c}{\textbf{5,000,000}} \\ 
\cline{2-3} \cline{5-6} \cline{8-9} \cline{11-12} 
& mean & median & & mean & median & & mean & median & & mean & median \\ \hline
10   & -0.0026 & -0.0026 &  & 0.0003 & 0.0007 &  & 0.7010 & 0.7210 &  & 5.9778 & 5.4425 \\ 
  20   & -0.0004 & -0.0003 &  & -0.0003 &  0.0000 &  & 0.5586 & 0.5295 &  & 4.4681 & 3.8385 \\ 
  100  & -0.0002 & -0.0002 &  & -0.0002 & -0.0002 &  & 0.8311 & 0.7720 &  & 6.3787 & 6.3190 \\ 
  500   & -0.0002 & -0.0002 &  & -0.0001 & -0.0001 &  & 4.6868 & 4.4250 &  & 19.9964 & 19.7895 \\ 
  1,000   & -0.0002 & -0.0002 &  & -0.0001 & -0.0001 &  & 14.1436 & 14.9040 &  & 42.1839 & 41.4760 \\
\hline\hline
\end{tabular}
\end{table}

\begin{table}[ht]
\centering
\scriptsize
\caption{This table considers the discrete simulation setting with $\mu=0.10$ and $\kappa=150$. The mean and median bias and run time are shown for the approximation based on the trapezium rule, the marginal approximation (in function of the number of boundaries) and the $k$-means approximation (in function of the number of clusters). Two data set sizes are considered, namely 500,000 and 5,000,000.}
\label{tab: mean-med, mu=0.1, conc=150}
\begin{tabular}{cccccccccccc}
\hline\hline
  & \multicolumn{5}{c}{bias} & & \multicolumn{5}{c}{run time (s)}\\ 
 \cline{2-6} \cline{8-12} 
\multirow{2}{*}{\textbf{\textit{k}-means}}  & \multicolumn{2}{c}{\textbf{500,000}} & & \multicolumn{2}{c}{\textbf{5,000,000}} & & \multicolumn{2}{c}{\textbf{500,000}} & & \multicolumn{2}{c}{\textbf{5,000,000}} \\ 
\cline{2-3} \cline{5-6} \cline{8-9} \cline{11-12} 
& mean & median & & mean & median & & mean & median & & mean & median \\ \hline
10  & -0.0068 & -0.0068 &  & -0.0024 & -0.0029 &  & 0.5268 & 0.4970 &  & 4.6623 & 4.6470 \\ 
  20   & 0.0001 & 0.0003 &  & -0.0002 & -0.0001 &  & 0.5864 & 0.5600 &  & 4.5954 & 3.9205 \\ 
  100   & -0.0001 & -0.0001 &  & -0.0001 & -0.0001 &  & 0.8301 & 0.7720 &  & 6.3045 & 6.2565 \\ 
  500   &  0.0000 &  0.0000 &  & -0.0001 & -0.0001 &  & 4.6777 & 4.4325 &  & 19.8896 & 19.6850 \\ 
  1,000   &  0.0000 &  0.0000 &  & -0.0001 & -0.0001 &  & 14.0217 & 13.1865 &  & 42.1063 & 41.3815 \\
\hline\hline
\end{tabular}
\end{table}

\begin{table}[ht]
\centering
\scriptsize
\caption{This table considers the discrete simulation setting with $\mu=0.25$ and $\kappa=50$. The mean and median bias and run time are shown for the approximation based on the trapezium rule, the marginal approximation (in function of the number of boundaries) and the $k$-means approximation (in function of the number of clusters). Two data set sizes are considered, namely 500,000 and 5,000,000.}
\label{tab: mean-med, mu=0.25, conc=50}
\begin{tabular}{cccccccccccc}
\hline\hline
  & \multicolumn{5}{c}{bias} & & \multicolumn{5}{c}{run time (s)}\\ 
 \cline{2-6} \cline{8-12} 
\multirow{2}{*}{\textbf{\textit{k}-means}}  & \multicolumn{2}{c}{\textbf{500,000}} & & \multicolumn{2}{c}{\textbf{5,000,000}} & & \multicolumn{2}{c}{\textbf{500,000}} & & \multicolumn{2}{c}{\textbf{5,000,000}} \\ 
\cline{2-3} \cline{5-6} \cline{8-9} \cline{11-12} 
& mean & median & & mean & median & & mean & median & & mean & median \\ \hline
10   & -0.0109 & -0.0136 &  & -0.0029 & -0.0022 &  & 0.5227 & 0.5020 &  & 4.6176 & 4.5790 \\ 
  20   & 0.0001 & 0.0003 &  & 0.0002 & 0.0003 &  & 0.6095 & 0.5700 &  & 4.6541 & 4.1635 \\ 
  100   & 0.0000 & 0.0000 &  & 0.0000 & 0.0000 &  & 0.8381 & 0.7950 &  & 6.3364 & 6.2920 \\ 
  500   & 0.0001 & 0.0000 &  & 0.0000 & 0.0000 &  & 5.1556 & 5.1110 &  & 19.8713 & 19.8500 \\ 
  1,000   & 0.0001 & 0.0000 &  & 0.0000 & 0.0000 &  & 16.2201 & 16.4855 &  & 40.8874 & 40.8250 \\
\hline\hline
\end{tabular}
\end{table}

\begin{table}[ht]
\centering
\scriptsize
\caption{This table considers the discrete simulation setting with $\mu=0.75$ and $\kappa=50$. The mean and median bias and run time are shown for the approximation based on the trapezium rule, the marginal approximation (in function of the number of boundaries) and the $k$-means approximation (in function of the number of clusters). Two data set sizes are considered, namely 500,000 and 5,000,000.}
\label{tab: mean-med, mu=0.75, conc=50}
\begin{tabular}{ccccccccccccc}
\hline\hline
  & \multicolumn{5}{c}{bias} & & \multicolumn{5}{c}{run time (s)}\\ 
 \cline{2-6} \cline{8-12} 
\multirow{2}{*}{\textbf{\textit{k}-means}}  & \multicolumn{2}{c}{\textbf{500,000}} & & \multicolumn{2}{c}{\textbf{5,000,000}} & & \multicolumn{2}{c}{\textbf{500,000}} & & \multicolumn{2}{c}{\textbf{5,000,000}} \\ 
\cline{2-3} \cline{5-6} \cline{8-9} \cline{11-12} 
& mean & median & & mean & median & & mean & median & & mean & median \\ \hline
10   & -0.0114 & -0.0152 &  & -0.0016 & -0.0008 &  & 0.5228 & 0.5080 &  & 4.6126 & 4.5490 \\ 
  20   &  0.0000 & 0.0003 &  & 0.0000 & 0.0000 &  & 0.6315 & 0.5995 &  & 4.7993 & 4.2885 \\ 
  100   & 0.0000 &  0.0000 &  &  0.0000 & 0.0000 &  & 0.8536 & 0.8070 &  & 6.4393 & 6.4040 \\ 
  500   & 0.0001 & 0.0001 &  & 0.0000 & 0.0000 &  & 5.1375 & 5.1690 &  & 19.7552 & 19.7420 \\ 
  1,000   & 0.0001 & 0.0001 &  & 0.0000 & 0.0000 &  & 15.8329 & 16.1250 &  & 41.2013 & 41.1345 \\
\hline\hline
\end{tabular}
\end{table}

\begin{table}[ht]
\centering
\scriptsize
\caption{This table considers the discrete simulation setting with $\mu=0.05$ and $\kappa=50$. The mean and median bias and run time are shown for the approximation based on the trapezium rule, the marginal approximation (in function of the number of boundaries) and the $k$-means approximation (in function of the number of clusters). Two data set sizes are considered, namely 500,000 and 5,000,000.}
\label{tab: mean-med, mu=0.05, conc=50}
\begin{tabular}{cccccccccccc}
\hline\hline
 & \multicolumn{5}{c}{bias} & & \multicolumn{5}{c}{run time (s)}\\ 
 \cline{2-6} \cline{8-12} 
\multirow{2}{*}{\textbf{\textit{k}-means}}  & \multicolumn{2}{c}{\textbf{500,000}} & & \multicolumn{2}{c}{\textbf{5,000,000}} & & \multicolumn{2}{c}{\textbf{500,000}} & & \multicolumn{2}{c}{\textbf{5,000,000}} \\ 
\cline{2-3} \cline{5-6} \cline{8-9} \cline{11-12} 
& mean & median & & mean & median & & mean & median & & mean & median \\ \hline
10  & 0.0076 & 0.0094 &  & 0.0005 & 0.0007 &  & 0.6477 & 0.6230 &  & 5.2933 & 4.8940 \\ 
  20   & 0.0000 & 0.0002 &  & -0.0005 & -0.0005 &  & 0.5676 & 0.5520 &  & 4.5119 & 3.8110 \\ 
  100   &  0.0000 &  0.0000 &  &  0.0000 & 0.0000 &  & 0.8150 & 0.7665 &  & 6.3971 & 6.3680 \\ 
  500   & 0.0000 & 0.0001 &  & 0.0000 & 0.0001 &  & 4.4340 & 4.1700 &  & 20.3406 & 20.3720 \\ 
  1,000   & 0.0000 & 0.0001 &  & 0.0001 & 0.0001 &  & 13.6279 & 12.7590 &  & 42.9405 & 42.6960 \\
\hline\hline
\end{tabular}
\end{table}

\begin{table}[ht]
\centering
\scriptsize
\caption{This table considers the discrete simulation setting with $\mu=0.95$ and $\kappa=50$. The mean and median bias and run time are shown for the approximation based on the trapezium rule, the marginal approximation (in function of the number of boundaries) and the $k$-means approximation (in function of the number of clusters). Two data set sizes are considered, namely 500,000 and 5,000,000.}
\label{tab: mean-med, mu=0.95, conc=50}
\begin{tabular}{cccccccccccc}
\hline\hline
 & \multicolumn{5}{c}{bias} & & \multicolumn{5}{c}{run time (s)}\\ 
 \cline{2-6} \cline{8-12} 
\multirow{2}{*}{\textbf{\textit{k}-means}}  & \multicolumn{2}{c}{\textbf{500,000}} & & \multicolumn{2}{c}{\textbf{5,000,000}} & & \multicolumn{2}{c}{\textbf{500,000}} & & \multicolumn{2}{c}{\textbf{5,000,000}} \\ 
\cline{2-3} \cline{5-6} \cline{8-9} \cline{11-12} 
& mean & median & & mean & median & & mean & median & & mean & median \\ \hline
10  & 0.0067 & 0.0085 &  & 0.0011 & 0.0019 &  & 0.6375 & 0.6085 &  & 5.1385 & 4.7980 \\ 
  20  & -0.0003 & -0.0001 &  & -0.0004 & -0.0004 &  & 0.5611 & 0.5555 &  & 4.4908 & 3.8060 \\ 
  100   & 0.0000 & 0.0000 &  & 0.0000 & 0.0000 &  & 0.8211 & 0.7680 &  & 6.3411 & 6.3105 \\ 
  500   & 0.0000 & 0.0001 &  & 0.0000 & 0.0001 &  & 4.4137 & 4.1560 &  & 20.1059 & 20.1765 \\ 
  1,000   & 0.0000 & 0.0001 &  & 0.0001 & 0.0001 &  & 13.6509 & 12.7190 &  & 42.9336 & 42.6280 \\
\hline\hline
\end{tabular}
\end{table}

\begin{table}[ht]
\centering
\scriptsize
\caption{This table considers the discrete simulation setting with $\mu=0.25$ and $\kappa=50$. The interquartile range and standard deviation over the estimates and over the run time are shown in function of the number of boundary values or clusters. This is done for the
$k$-means approximation, the marginal approximation and the one based on the trapezium rule. Two data set sizes are considered, namely 500,000 and 5,000,000.}
\label{tab: sd-IQR, mu=0.25, conc=50}
\begin{tabular}{cccccccccccc}
\hline\hline
 &  \multicolumn{5}{c}{$\hat{C}$} & & \multicolumn{5}{c}{run time (s)}\\ 
 \cline{2-6} \cline{8-12} 
\multirow{2}{*}{\textbf{\textit{k}-means}} & \multicolumn{2}{c}{\textbf{500,000}} & & \multicolumn{2}{c}{\textbf{5,000,000}} & & \multicolumn{2}{c}{\textbf{500,000}} & & \multicolumn{2}{c}{\textbf{5,000,000}} \\ 
\cline{2-3} \cline{5-6} \cline{8-9} \cline{11-12} 
&  $\sigma$ & IQR & & $\sigma$ & IQR & & $\sigma$ & IQR & & $\sigma$ & IQR \\ \hline
10   & 0.0198 & 0.0307 &  & 0.0196 & 0.0311 &  & 0.1284 & 0.1597 &  & 1.0836 & 1.4550 \\ 
  20   & 0.0051 & 0.0063 &  & 0.0044 & 0.0058 &  & 0.1896 & 0.2755 &  & 1.3292 & 1.6480 \\ 
  100   & 0.0010 & 0.0013 &  & 0.0004 & 0.0006 &  & 0.1489 & 0.1880 &  & 0.2969 & 0.1203 \\ 
  500   & 0.0009 & 0.0013 &  & 0.0003 & 0.0004 &  & 0.6278 & 0.9455 &  & 0.2888 & 0.1842 \\ 
  1,000   & 0.0009 & 0.0012 &  & 0.0003 & 0.0004 &  & 1.1688 & 1.6800 &  & 0.6113 & 0.2740 \\
\hline\hline
\end{tabular}
\end{table}

\begin{table}[ht]
\centering
\scriptsize
\caption{This table considers the discrete simulation setting with $\mu=0.50$ and $\kappa=50$. The interquartile range and standard deviation over the estimates and over the run time are shown in function of the number of boundary values or clusters. This is done for the
$k$-means approximation, the marginal approximation and the one based on the trapezium rule. Two data set sizes are considered, namely 500,000 and 5,000,000.}
\label{tab: sd-IQR, mu=0.5, conc=50}
\begin{tabular}{cccccccccccc}
\hline\hline
 &  \multicolumn{5}{c}{$\hat{C}$} & & \multicolumn{5}{c}{run time (s)}\\ 
 \cline{2-6} \cline{8-12} 
\multirow{2}{*}{\textbf{\textit{k}-means}}  & \multicolumn{2}{c}{\textbf{500,000}} & & \multicolumn{2}{c}{\textbf{5,000,000}} & & \multicolumn{2}{c}{\textbf{500,000}} & & \multicolumn{2}{c}{\textbf{5,000,000}} \\ 
\cline{2-3} \cline{5-6} \cline{8-9} \cline{11-12} 
& $\sigma$ & IQR & & $\sigma$ & IQR & & $\sigma$ & IQR & & $\sigma$ & IQR \\ \hline
10  & 0.0191 & 0.0267 &  & 0.0193 & 0.0283 &  & 0.1101 & 0.1453 &  & 0.9143 & 1.1840 \\ 
  20   & 0.0051 & 0.0063 &  & 0.0045 & 0.0058 &  & 0.1812 & 0.2725 &  & 1.2505 & 1.5353 \\ 
  100   & 0.0008 & 0.0011 &  & 0.0004 & 0.0006 &  & 0.1479 & 0.2130 &  & 0.2409 & 0.1015 \\ 
  500 & 0.0008 & 0.0010 &  & 0.0003 & 0.0003 &  & 0.5301 & 0.6510 &  & 0.1321 & 0.1782 \\ 
  1,000   & 0.0008 & 0.0010 &  & 0.0003 & 0.0003 &  & 1.9919 & 3.0160 &  & 0.2073 & 0.2715 \\
\hline\hline
\end{tabular}
\end{table}

\begin{table}[ht]
\centering
\scriptsize
\caption{This table considers the discrete simulation setting with $\mu=0.10$ and $\kappa=15$. The interquartile range and standard deviation over the estimates and over the run time are shown in function of the number of boundary values or clusters. This is done for the
$k$-means approximation, the marginal approximation and the one based on the trapezium rule. Two data set sizes are considered, namely 500,000 and 5,000,000.}
\label{tab: sd-IQR, mu=0.1, conc=15}
\begin{tabular}{cccccccccccc}
\hline\hline
 & \multicolumn{5}{c}{$\hat{C}$} & & \multicolumn{5}{c}{run time (s)}\\ 
 \cline{2-6} \cline{8-12} 
\multirow{2}{*}{\textbf{\textit{k}-means}}  & \multicolumn{2}{c}{\textbf{500,000}} & & \multicolumn{2}{c}{\textbf{5,000,000}} & & \multicolumn{2}{c}{\textbf{500,000}} & & \multicolumn{2}{c}{\textbf{5,000,000}} \\ 
\cline{2-3} \cline{5-6} \cline{8-9} \cline{11-12} 
&  $\sigma$ & IQR & & $\sigma$ & IQR & & $\sigma$ & IQR & & $\sigma$ & IQR \\ \hline
10  & 0.0116 & 0.0157 &  & 0.0123 & 0.0178 &  & 0.2456 & 0.3935 &  & 2.3443 & 4.0407 \\ 
  20   & 0.0037 & 0.0050 &  & 0.0036 & 0.0048 &  & 0.1951 & 0.2400 &  & 1.3723 & 1.2717 \\ 
  100   & 0.0011 & 0.0015 &  & 0.0005 & 0.0006 &  & 0.1530 & 0.1140 &  & 0.2567 & 0.1832 \\ 
  500   & 0.0011 & 0.0015 &  & 0.0004 & 0.0005 &  & 0.5078 & 1.0070 &  & 0.5364 & 0.3355 \\ 
  1,000  & 0.0011 & 0.0015 &  & 0.0004 & 0.0005 &  & 1.1389 & 2.0965 &  & 1.2206 & 2.3445 \\
\hline\hline
\end{tabular}
\end{table}

\begin{table}[ht]
\centering
\scriptsize
\caption{This table considers the discrete simulation setting with $\mu=0.10$ and $\kappa=150$. The interquartile range and standard deviation over the estimates and over the run time are shown in function of the number of boundary values or clusters. This is done for the
$k$-means approximation, the marginal approximation and the one based on the trapezium rule. Two data set sizes are considered, namely 500,000 and 5,000,000.}
\label{tab: sd-IQR, mu=0.1, conc=150}
\begin{tabular}{cccccccccccc}
\hline\hline
  & \multicolumn{5}{c}{$\hat{C}$} & & \multicolumn{5}{c}{run time (s)}\\ 
 \cline{2-6} \cline{8-12} 
\multirow{2}{*}{\textbf{\textit{k}-means}} & \multicolumn{2}{c}{\textbf{500,000}} & & \multicolumn{2}{c}{\textbf{5,000,000}} & & \multicolumn{2}{c}{\textbf{500,000}} & & \multicolumn{2}{c}{\textbf{5,000,000}} \\ 
\cline{2-3} \cline{5-6} \cline{8-9} \cline{11-12} 
&  $\sigma$ & IQR & & $\sigma$ & IQR & & $\sigma$ & IQR & & $\sigma$ & IQR \\ \hline
10   & 0.0170 & 0.0241 &  & 0.0180 & 0.0237 &  & 0.1543 & 0.1932 &  & 1.2438 & 1.8035 \\ 
  20   & 0.0050 & 0.0062 &  & 0.0045 & 0.0059 &  & 0.2057 & 0.3150 &  & 1.4313 & 1.7775 \\ 
  100   & 0.0014 & 0.0019 &  & 0.0006 & 0.0007 &  & 0.1503 & 0.1068 &  & 0.2420 & 0.1345 \\ 
  500   & 0.0013 & 0.0018 &  & 0.0004 & 0.0006 &  & 0.5007 & 0.2217 &  & 0.5241 & 0.2482 \\ 
  1,000   & 0.0013 & 0.0018 &  & 0.0004 & 0.0006 &  & 1.1029 & 2.1043 &  & 1.2087 & 2.3322 \\
\hline\hline
\end{tabular}
\end{table}

\begin{table}[ht]
\centering
\scriptsize
\caption{This table considers the discrete simulation setting with $\mu=0.75$ and $\kappa=50$. The interquartile range and standard deviation over the estimates and over the run time are shown in function of the number of boundary values or clusters. This is done for the
$k$-means approximation, the marginal approximation and the one based on the trapezium rule. Two data set sizes are considered, namely 500,000 and 5,000,000.}
\label{tab: sd-IQR, mu=0.75, conc=50}
\begin{tabular}{cccccccccccc}
\hline\hline
 &  \multicolumn{5}{c}{$\hat{C}$} & & \multicolumn{5}{c}{run time (s)}\\ 
 \cline{2-6} \cline{8-12} 
\multirow{2}{*}{\textbf{\textit{k}-means}}  & \multicolumn{2}{c}{\textbf{500,000}} & & \multicolumn{2}{c}{\textbf{5,000,000}} & & \multicolumn{2}{c}{\textbf{500,000}} & & \multicolumn{2}{c}{\textbf{5,000,000}} \\ 
\cline{2-3} \cline{5-6} \cline{8-9} \cline{11-12} 
&  $\sigma$ & IQR & & $\sigma$ & IQR & & $\sigma$ & IQR & & $\sigma$ & IQR \\ \hline
10   & 0.0196 & 0.0300 &  & 0.0188 & 0.0273 &  & 0.1293 & 0.1562 &  & 1.1105 & 1.5400 \\ 
  20   & 0.0052 & 0.0065 &  & 0.0045 & 0.0059 &  & 0.1964 & 0.2985 &  & 1.3582 & 1.8228 \\ 
  100   & 0.0010 & 0.0013 &  & 0.0004 & 0.0006 &  & 0.1546 & 0.1905 &  & 0.2698 & 0.1200 \\ 
  500   & 0.0009 & 0.0012 &  & 0.0003 & 0.0004 &  & 0.6318 & 0.9642 &  & 0.2204 & 0.1755 \\ 
  1,000   & 0.0009 & 0.0013 &  & 0.0003 & 0.0004 &  & 1.1310 & 1.6823 &  & 0.6470 & 0.3215 \\
\hline\hline
\end{tabular}
\end{table}

\begin{table}[ht]
\centering
\scriptsize
\caption{This table considers the discrete simulation setting with $\mu=0.05$ and $\kappa=50$. The interquartile range and standard deviation over the estimates and over the run time are shown in function of the number of boundary values or clusters. This is done for the
$k$-means approximation, the marginal approximation and the one based on the trapezium rule. Two data set sizes are considered, namely 500,000 and 5,000,000.}
\label{tab: sd-IQR, mu=0.05, conc=50}
\begin{tabular}{cccccccccccc}
\hline\hline
 &  \multicolumn{5}{c}{$\hat{C}$} & & \multicolumn{5}{c}{run time (s)}\\ 
 \cline{2-6} \cline{8-12} 
\multirow{2}{*}{\textbf{\textit{k}-means}}  & \multicolumn{2}{c}{\textbf{500,000}} & & \multicolumn{2}{c}{\textbf{5,000,000}}  && \multicolumn{2}{c}{\textbf{500,000}} & & \multicolumn{2}{c}{\textbf{5,000,000}} \\ 
\cline{2-3} \cline{5-6} \cline{8-9} \cline{11-12} 
&  $\sigma$ & IQR & & $\sigma$ & IQR & & $\sigma$ & IQR & & $\sigma$ & IQR \\ \hline
10  & 0.0154 & 0.0222 &  & 0.0140 & 0.0202 &  & 0.2448 & 0.3895 &  & 2.0553 & 3.1440 \\ 
  20   & 0.0043 & 0.0054 &  & 0.0040 & 0.0052 &  & 0.2009 & 0.2728 &  & 1.4548 & 1.8207 \\ 
  100   & 0.0017 & 0.0024 &  & 0.0006 & 0.0008 &  & 0.1435 & 0.0620 &  & 0.2681 & 0.2150 \\ 
  500   & 0.0017 & 0.0022 &  & 0.0005 & 0.0007 &  & 0.5085 & 0.1390 &  & 0.4325 & 0.5770 \\ 
  1,000   & 0.0017 & 0.0022 &  & 0.0005 & 0.0007 &  & 1.1421 & 2.2143 &  & 1.5034 & 1.9625 \\
\hline\hline
\end{tabular}
\end{table}

\begin{table}[ht]
\centering
\scriptsize
\caption{This table considers the discrete simulation setting with $\mu=0.95$ and $\kappa=50$. The interquartile range and standard deviation over the estimates and over the run time are shown in function of the number of boundary values or clusters. This is done for the $k$-means approximation, the marginal approximation and the one based on the trapezium rule. Two data set sizes are considered, namely 500,000 and 5,000,000.}
\label{tab: sd-IQR, mu=0.95, conc=50}
\begin{tabular}{cccccccccccc}
\hline\hline
 &  \multicolumn{5}{c}{$\hat{C}$} & & \multicolumn{5}{c}{run time (s)}\\ 
 \cline{2-6} \cline{8-12} 
\multirow{2}{*}{\textbf{\textit{k}-means}}  & \multicolumn{2}{c}{\textbf{500,000}} & & \multicolumn{2}{c}{\textbf{5,000,000}} & & \multicolumn{2}{c}{\textbf{500,000}} & & \multicolumn{2}{c}{\textbf{5,000,000}} \\ 
\cline{2-3} \cline{5-6} \cline{8-9} \cline{11-12} 
&  $\sigma$ & IQR & & $\sigma$ & IQR & & $\sigma$ & IQR & & $\sigma$ & IQR \\ \hline
10   & 0.0152 & 0.0227 &  & 0.0142 & 0.0202 &  & 0.2395 & 0.3865 &  & 1.9812 & 2.5878 \\ 
  20   & 0.0045 & 0.0058 &  & 0.0040 & 0.0053 &  & 0.1918 & 0.2650 &  & 1.4006 & 1.8860 \\ 
  100   & 0.0017 & 0.0022 &  & 0.0006 & 0.0009 &  & 0.1499 & 0.0633 &  & 0.2625 & 0.2033 \\ 
  500   & 0.0017 & 0.0022 &  & 0.0005 & 0.0007 &  & 0.5091 & 0.1062 &  & 0.4072 & 0.6152 \\ 
  1,000   & 0.0017 & 0.0022 &  & 0.0005 & 0.0007 &  & 1.1652 & 2.2292 &  & 1.4979 & 2.0735 \\
\hline\hline
\end{tabular}
\end{table}

\begin{table}[ht]
\centering
\scriptsize
\caption{This table considers the discrete simulation setting with $\mu=0.25$ and $\kappa=50$. It shows the mean and median bias and run time, but also the interquartile range and standard deviation over the estimates and over the run time. This all is shown in function of the number of boundary values or clusters, for the $k$-means approximation, the marginal approximation and the one based on the trapezium rule. The size of the considered data sets is 50,000,000.}
\label{tab: sd-IQR, mu=0.95, conc=50, N=50,000,000}
\begin{tabular}{cccccccccccc}
\hline\hline
 &  \multicolumn{4}{c}{$\hat{C}$} & & \multicolumn{4}{c}{run time (s)}\\ 
 \cline{2-5} \cline{7-10} 
 \textbf{\textit{k}-means}   & mean & median & $\sigma$ & IQR & & mean & median & $\sigma$ & IQR \\ \hline
 10   & -0.0007 & 0.0009 &  0.0173 & 0.0242 &  & 46.7064 & 46.8180   & 9.4009 & 15.1730 \\ 
  20   & 0.0000 & -0.0002 &  0.0043 & 0.0058 &  & 43.8656 & 39.0630   & 9.6861 & 7.8725 \\ 
  100   & 0.0000 & 0.0000 &  0.0004 & 0.0005 &  & 64.0477 & 64.0680   & 0.8296 & 1.1298 \\ 
  500   & 0.0000 & 0.0000 &   0.0001 & 0.0001 &  & 177.8738 & 177.9935 &   0.9781 & 1.3748 \\ 
  1,000   & 0.0000 & 0.0000 &   0.0001 & 0.0001 &  & 327.5612 & 327.4250 &   1.5166 & 1.9970 \\ 
 \hline
\textbf{marginal}   & mean & median & $\sigma$ & IQR & & mean & median & $\sigma$ & IQR \\ \hline
 10  & 0.0083 & 0.0083 & 0.0001 & 0.0001 &  & 14.9257 & 14.9190   & 0.0718 & 0.0890 \\ 
  20   & 0.0044 & 0.0044   & 0.0001 & 0.0001 &  & 14.9681 & 14.9650   & 0.0737 & 0.0930 \\ 
  100   & 0.0009 & 0.0009   & 0.0001 & 0.0001 &  & 15.8595 & 15.8545   & 0.0980 & 0.1373 \\ 
  500  & 0.0002 & 0.0002   & 0.0001 & 0.0001 &  & 20.2807 & 20.2785   & 0.0829 & 0.1110 \\ 
  1,000   & 0.0001 & 0.0001  & 0.0001 & 0.0001 &  & 25.8803 & 25.8785  & 0.0981 & 0.1283 \\ 
  \hline
  \textbf{trapezium}   & mean & median & $\sigma$ & IQR & & mean & median & $\sigma$ & IQR \\ \hline
  & 0.0000 & 0.0000 & 0.0001 & 0.0001 &  & 29.3463 & 29.3615  & 0.1338 & 0.1775 \\ 
\hline\hline
\end{tabular}
\end{table}
\FloatBarrier

\section*{Appendix C.}
In this appendix, the results of the simulation study in the continuous setting are shown.

\begin{table}[ht]
\centering
\scriptsize
\caption{This table considers the continuous simulation setting with $\rho=0$ and $\nu=0$. The mean and median bias and run time are shown in function of the number clusters for the $k$-means approximation and the number of boundary values for the marginal approximation. Two data set sizes are considered, namely 500,000 and 5,000,000.}
\label{tab: bias-time,  rho=0 and nu=0}
\begin{tabular}{cccccccccccc}
\hline\hline
 & \multicolumn{5}{c}{bias} & & \multicolumn{5}{c}{run time (s)}\\ 
 \cline{2-6} \cline{8-12} 
\multirow{2}{*}{\textbf{\textit{k}-means}}  & \multicolumn{2}{c}{\textbf{500,000}} & & \multicolumn{2}{c}{\textbf{5,000,000}} & & \multicolumn{2}{c}{\textbf{500,000}} & & \multicolumn{2}{c}{\textbf{5,000,000}} \\ 
\cline{2-3} \cline{5-6} \cline{8-9} \cline{11-12} 
&  mean & median & & mean & median & & mean & median & & mean & median \\ \hline
10   & -0.0024 & -0.0029 &  & 0.0022 & 0.0038 &  & 1.0263 & 0.9195 &  & 7.0400 & 5.5005 \\ 
  20   & 0.0005 & 0.0001 &  & -0.0003 & -0.0012 &  & 1.1580 & 0.9765 &  & 7.3263 & 5.9520 \\ 
  100  & 0.0000 & 0.0001 &  & 0.0000 & 0.0000 &  & 2.2384 & 1.6940 &  & 11.3656 & 10.3800 \\ 
 500   &  0.0000 &  0.0000 &  &  0.0000 &  0.0000 &  & 15.2763 & 15.6920 &  & 26.8637 & 26.4390 \\ 
  1,000   &  0.0000 &  0.0000 &  &  0.0000 &  0.0000 &  & 34.3116 & 35.2730 &  & 46.4865 & 46.1975 \\ 
\hline
\multirow{2}{*}{\textbf{marginal}} & \multicolumn{2}{c}{\textbf{500,000}} & & \multicolumn{2}{c}{\textbf{5,000,000}} & & \multicolumn{2}{c}{\textbf{500,000}} & & \multicolumn{2}{c}{\textbf{5,000,000}} \\ 
\cline{2-3} \cline{5-6} \cline{8-9} \cline{11-12} 
&  mean & median & & mean & median & & mean & median & & mean & median \\ \hline
  10  & 0.0000 & 0.0000 &  & 0.0000 & 0.0000 &  & 2.0515 & 2.0520 &  & 26.7363 & 26.6895 \\ 
  20  & 0.0000 & 0.0000 &  & 0.0000 & 0.0000 &  & 3.6855 & 3.6870 &  & 49.0309 & 48.9470 \\ 
  100  & 0.0000 & 0.0000 &  & 0.0000 & 0.0000 &  & 18.6691 & 18.6560 &  & 250.4291 & 248.8755 \\ 
\hline\hline
\end{tabular}
\end{table}

\begin{table}[ht]
\centering
\scriptsize
\caption{This table considers the continuous simulation setting with $\rho=0$ and $\nu=0.3583$. The mean and median bias and run time are shown in function of the number clusters for the $k$-means approximation and the number of boundary values for the marginal approximation. Two data set sizes are considered, namely 500,000 and 5,000,000.}
\label{tab: bias-time, rho=0 and nu=0.3583}
\begin{tabular}{cccccccccccc}
\hline\hline
 & \multicolumn{5}{c}{bias} & & \multicolumn{5}{c}{run time (s)}\\ 
 \cline{2-6} \cline{8-12} 
\multirow{2}{*}{\textbf{\textit{k}-means}}  & \multicolumn{2}{c}{\textbf{500,000}} & & \multicolumn{2}{c}{\textbf{5,000,000}} & & \multicolumn{2}{c}{\textbf{500,000}} & & \multicolumn{2}{c}{\textbf{5,000,000}} \\ 
\cline{2-3} \cline{5-6} \cline{8-9} \cline{11-12} 
&  mean & median & & mean & median & & mean & median & & mean & median \\ \hline
10 &   -0.0024 & -0.0021 &  & 0.0005 & 0.0018 &  & 1.0285 & 0.9245 &  & 7.0952 & 5.5435 \\ 
  20 & 0.0001 & 0.0001 &  & -0.0001 &  0.0000 &  & 1.1580 & 0.9745 &  & 7.3927 & 6.0075 \\ 
  100   &  0.0000 &  0.0000 &  & -0.0001 & -0.0001 &  & 2.2287 & 1.6840 &  & 11.2822 & 10.2935 \\ 
  500   &  0.0000 & -0.0001 &  & -0.0001 & -0.0001 &  & 15.1196 & 15.5025 &  & 26.7356 & 26.3135 \\ 
  1,000  & -0.0001 & -0.0001 &  & -0.0001 & -0.0001 &  & 33.6731 & 34.6105 &  & 45.9020 & 45.6100 \\ 
\hline
\multirow{2}{*}{\textbf{marginal}} & \multicolumn{2}{c}{\textbf{500,000}} & & \multicolumn{2}{c}{\textbf{5,000,000}} & & \multicolumn{2}{c}{\textbf{500,000}} & & \multicolumn{2}{c}{\textbf{5,000,000}} \\ 
\cline{2-3} \cline{5-6} \cline{8-9} \cline{11-12} 
&  mean & median & & mean & median & & mean & median & & mean & median \\ \hline
  10 & -0.0001 & -0.0001 &  & -0.0001 & -0.0001 &  & 2.0994 & 2.1000 &  & 26.7484 & 26.7860 \\ 
  20 & -0.0001 & -0.0001 &  & -0.0001 & -0.0001 &  & 3.8516 & 3.8520 &  & 48.9944 & 49.0825 \\ 
  100   & -0.0001 & -0.0001 &  & -0.0001 & -0.0001 &  & 84.3169 & 84.3170 &  & 317.8971 & 315.8615 \\ 
\hline\hline
\end{tabular}
\end{table}

\begin{table}[ht]
\centering
\scriptsize
\caption{This table considers the continuous simulation setting with $\rho=0$ and $\nu=0.7416$. The mean and median bias and run time are shown in function of the number clusters for the $k$-means approximation and the number of boundary values for the marginal approximation. Two data set sizes are considered, namely 500,000 and 5,000,000.}
\label{tab: bias-time, rho=0 and nu=0.7416}
\begin{tabular}{cccccccccccc}
\hline\hline
 & \multicolumn{5}{c}{bias} & & \multicolumn{5}{c}{run time (s)}\\ 
 \cline{2-6} \cline{8-12} 
\multirow{2}{*}{\textbf{\textit{k}-means}}  & \multicolumn{2}{c}{\textbf{500,000}} & & \multicolumn{2}{c}{\textbf{5,000,000}} & & \multicolumn{2}{c}{\textbf{500,000}} & & \multicolumn{2}{c}{\textbf{5,000,000}} \\ 
\cline{2-3} \cline{5-6} \cline{8-9} \cline{11-12} 
&  mean & median & & mean & median & & mean & median & & mean & median \\ \hline
10 &   -0.0022 & -0.0014 &  & 0.0003 & 0.0000 &  & 1.0246 & 0.9170 &  & 7.0611 & 5.5180 \\ 
  20 & 0.0008 & 0.0002 &  & 0.0003 & 0.0011 &  & 1.1522 & 0.9705 &  & 7.3847 & 6.0020 \\ 
  100   & -0.0001 & -0.0001 &  & 0.0001 & 0.0000 &  & 2.2238 & 1.6785 &  & 11.2907 & 10.2975 \\ 
  500   &  0.0000 & -0.0001 &  & -0.0001 & -0.0001 &  & 14.9875 & 15.4265 &  & 26.5436 & 26.1240 \\ 
  1,000  & -0.0001 & -0.0001 &  & -0.0001 & -0.0001 &  & 33.0831 & 34.0690 &  & 45.2205 & 44.9320 \\ 
\hline
\multirow{2}{*}{\textbf{marginal}} & \multicolumn{2}{c}{\textbf{500,000}} & & \multicolumn{2}{c}{\textbf{5,000,000}} & & \multicolumn{2}{c}{\textbf{500,000}} & & \multicolumn{2}{c}{\textbf{5,000,000}} \\ 
\cline{2-3} \cline{5-6} \cline{8-9} \cline{11-12} 
&  mean & median & & mean & median & & mean & median & & mean & median \\ \hline
10 &   -0.0001 & -0.0001 &  & -0.0001 & -0.0001 &  & 2.0833 & 2.0880 &  & 26.6854 & 26.7050 \\ 
  20 & -0.0001 & -0.0001 &  & -0.0001 & -0.0001 &  & 3.7804 & 3.7700 &  & 49.1028 & 48.9365 \\ 
  100   & -0.0001 & -0.0001 &  & -0.0001 & -0.0001 &  & 67.8759 & 67.9005 &  & 300.6183 & 298.5055 \\ 
\hline\hline
\end{tabular}
\end{table}

\begin{table}[ht]
\centering
\scriptsize
\caption{This table considers the continuous simulation setting with $\rho=0.25$ and $\nu=0$. The mean and median bias and run time are shown in function of the number clusters for the $k$-means approximation and the number of boundary values for the marginal approximation. Two data set sizes are considered, namely 500,000 and 5,000,000.}
\label{tab: bias-time,  rho=0.25 and nu=0}
\begin{tabular}{cccccccccccc}
\hline\hline
 & \multicolumn{5}{c}{bias} & & \multicolumn{5}{c}{run time (s)}\\ 
 \cline{2-6} \cline{8-12} 
\multirow{2}{*}{\textbf{\textit{k}-means}}  & \multicolumn{2}{c}{\textbf{500,000}} & & \multicolumn{2}{c}{\textbf{5,000,000}} & & \multicolumn{2}{c}{\textbf{500,000}} & & \multicolumn{2}{c}{\textbf{5,000,000}} \\ 
\cline{2-3} \cline{5-6} \cline{8-9} \cline{11-12} 
&  mean & median & & mean & median & & mean & median & & mean & median \\ \hline
10 &   0.0142 & 0.0174 &  & 0.0219 & 0.0239 &  & 1.0310 & 0.8880 &  & 6.8763 & 5.4495 \\ 
  20 & 0.0090 & 0.0095 &  & 0.0113 & 0.0108 &  & 1.1825 & 0.9810 &  & 7.4700 & 6.0335 \\ 
  100   & 0.0021 & 0.0021 &  & 0.0022 & 0.0022 &  & 2.2628 & 1.6840 &  & 11.4107 & 10.3405 \\ 
  500   & 0.0004 & 0.0005 &  & 0.0004 & 0.0004 &  & 14.5521 & 14.1795 &  & 26.7214 & 26.3120 \\ 
  1,000  & 0.0002 & 0.0003 &  & 0.0002 & 0.0002 &  & 33.7758 & 34.9045 &  & 46.2797 & 46.0320 \\ 
\hline
\multirow{2}{*}{\textbf{marginal}} & \multicolumn{2}{c}{\textbf{500,000}} & & \multicolumn{2}{c}{\textbf{5,000,000}} & & \multicolumn{2}{c}{\textbf{500,000}} & & \multicolumn{2}{c}{\textbf{5,000,000}} \\ 
\cline{2-3} \cline{5-6} \cline{8-9} \cline{11-12} 
&  mean & median & & mean & median & & mean & median & & mean & median \\ \hline
  10  & 0.0155 & 0.0155 &  & 0.0155 & 0.0155 &  & 2.0485 & 2.0505 &  & 26.8095 & 26.8445 \\ 
  20  & 0.0080 & 0.0080 &  & 0.0080 & 0.0079 &  & 3.6768 & 3.6810 &  & 49.1431 & 49.0935 \\ 
  100  & 0.0016 & 0.0017 &  & 0.0016 & 0.0016 &  & 18.6833 & 18.6850 &  & 250.5547 & 249.9060 \\ 
\hline\hline
\end{tabular}
\end{table}

\begin{table}[ht]
\centering
\scriptsize
\caption{This table considers the continuous simulation setting with $\rho=0.25$ and $\nu=0.7416$. The mean and median bias and run time are shown in function of the number clusters for the $k$-means approximation and the number of boundary values for the marginal approximation. Two data set sizes are considered, namely 500,000 and 5,000,000.}
\label{tab: bias-time, rho=0.25 and nu=0.7416}
\begin{tabular}{cccccccccccc}
\hline\hline
 & \multicolumn{5}{c}{bias} & & \multicolumn{5}{c}{run time (s)}\\ 
 \cline{2-6} \cline{8-12} 
\multirow{2}{*}{\textbf{\textit{k}-means}}  & \multicolumn{2}{c}{\textbf{500,000}} & & \multicolumn{2}{c}{\textbf{5,000,000}} & & \multicolumn{2}{c}{\textbf{500,000}} & & \multicolumn{2}{c}{\textbf{5,000,000}} \\ 
\cline{2-3} \cline{5-6} \cline{8-9} \cline{11-12} 
&  mean & median & & mean & median & & mean & median & & mean & median \\ \hline
10 &   0.0093 & 0.0095 &  & 0.0172 & 0.0168 &  & 1.0316 & 0.8890 &  & 6.7590 & 5.3490 \\ 
  20 & 0.0059 & 0.0065 &  & 0.0089 & 0.0084 &  & 1.1796 & 0.9790 &  & 7.3514 & 5.9230 \\ 
  100   & 0.0020 & 0.0020 &  & 0.0020 & 0.0019 &  & 2.2526 & 1.6725 &  & 11.2988 & 10.2390 \\ 
  500  & 0.0005 & 0.0005 &  & 0.0004 & 0.0004 &  & 14.3184 & 13.9990 &  & 26.3693 & 25.9665 \\ 
  1,000  & 0.0003 & 0.0003 &  & 0.0002 & 0.0002 &  & 32.4910 & 33.7730 &  & 44.8315 & 44.5885 \\ 
\hline
\multirow{2}{*}{\textbf{marginal}} & \multicolumn{2}{c}{\textbf{500,000}} & & \multicolumn{2}{c}{\textbf{5,000,000}} & & \multicolumn{2}{c}{\textbf{500,000}} & & \multicolumn{2}{c}{\textbf{5,000,000}} \\ 
\cline{2-3} \cline{5-6} \cline{8-9} \cline{11-12} 
&  mean & median & & mean & median & & mean & median & & mean & median \\ \hline
  10 & 0.0276 & 0.0276 &  & 0.0276 & 0.0276 &  & 2.0951 & 2.0935 &  & 26.6326 & 26.6720 \\ 
  20 & 0.0144 & 0.0144 &  & 0.0143 & 0.0143 &  & 3.8069 & 3.8020 &  & 49.1097 & 49.2280 \\ 
  100   & 0.0032 & 0.0032 &  & 0.0032 & 0.0032 &  & 68.0152 & 68.0225 &  & 298.9286 & 297.9795 \\ 
\hline\hline
\end{tabular}
\end{table}

\begin{table}[ht]
\centering
\scriptsize
\caption{This table considers the continuous simulation setting with $\rho=0.5$ and $\nu=0$. The mean and median bias and run time are shown in function of the number clusters for the $k$-means approximation and the number of boundary values for the marginal approximation. Two data set sizes are considered, namely 500,000 and 5,000,000.}
\label{tab: bias-time, rho=0.5 and nu=0}
\begin{tabular}{cccccccccccc}
\hline\hline
 & \multicolumn{5}{c}{bias} & & \multicolumn{5}{c}{run time (s)}\\ 
 \cline{2-6} \cline{8-12} 
\multirow{2}{*}{\textbf{\textit{k}-means}}  & \multicolumn{2}{c}{\textbf{500,000}} & & \multicolumn{2}{c}{\textbf{5,000,000}} & & \multicolumn{2}{c}{\textbf{500,000}} & & \multicolumn{2}{c}{\textbf{5,000,000}} \\ 
\cline{2-3} \cline{5-6} \cline{8-9} \cline{11-12} 
&  mean & median & & mean & median & & mean & median & & mean & median \\ \hline
10 &  0.0451 & 0.0426 &  & 0.0444 & 0.0426 &  & 1.3275 & 1.3760 &  & 11.0675 & 8.4615 \\ 
  20 & 0.0223 & 0.0216 &  & 0.0225 & 0.0228 &  & 1.2339 & 0.9960 &  & 7.6307 & 5.9180 \\ 
  100 & 0.0041 & 0.0043 &  & 0.0044 & 0.0044 &  & 2.3798 & 1.6730 &  & 11.2259 & 10.1895 \\ 
  500 & 0.0009 & 0.0008 &  & 0.0008 & 0.0008 &  & 13.9066 & 14.0420 &  & 26.2616 & 26.0010 \\ 
  1,000 & 0.0004 & 0.0004 &  & 0.0004 & 0.0004 &  & 33.5897 & 34.7080 &  & 45.4601 & 45.3240 \\ 
\hline
\multirow{2}{*}{\textbf{marginal}} & \multicolumn{2}{c}{\textbf{500,000}} & & \multicolumn{2}{c}{\textbf{5,000,000}} & & \multicolumn{2}{c}{\textbf{500,000}} & & \multicolumn{2}{c}{\textbf{5,000,000}} \\ 
\cline{2-3} \cline{5-6} \cline{8-9} \cline{11-12} 
&  mean & median & & mean & median & & mean & median & & mean & median \\ \hline
  10  & 0.0314 & 0.0314 &  & 0.0314 & 0.0314 &  & 2.3976 & 2.3970 &  & 26.8700 & 26.8700 \\ 
  20  & 0.0162 & 0.0162 &  & 0.0162 & 0.0162 &  & 4.3068 & 4.3100 &  & 49.2660 & 49.0730 \\ 
  100  & 0.0033 & 0.0033 &  & 0.0033 & 0.0033 &  & 21.5017 & 21.5230 &  & 249.1292 & 249.2600 \\ 
\hline\hline
\end{tabular}
\end{table}

\begin{table}[ht]
\centering
\scriptsize
\caption{This table considers the continuous simulation setting with $\rho=0.5$ and $\nu=0.3583$. The mean and median bias and run time are shown in function of the number clusters for the $k$-means approximation and the number of boundary values for the marginal approximation. Two data set sizes are considered, namely 500,000 and 5,000,000.}
\label{tab: bias-time, rho=0.5 and nu=0.3583}
\begin{tabular}{cccccccccccc}
\hline\hline
 & \multicolumn{5}{c}{bias} & & \multicolumn{5}{c}{run time (s)}\\ 
 \cline{2-6} \cline{8-12} 
\multirow{2}{*}{\textbf{\textit{k}-means}}  & \multicolumn{2}{c}{\textbf{500,000}} & & \multicolumn{2}{c}{\textbf{5,000,000}} & & \multicolumn{2}{c}{\textbf{500,000}} & & \multicolumn{2}{c}{\textbf{5,000,000}} \\ 
\cline{2-3} \cline{5-6} \cline{8-9} \cline{11-12} 
&  mean & median & & mean & median & & mean & median & & mean & median \\ \hline
10 &  0.0562 & 0.0745 &  & 0.0482 & 0.0513 &  & 1.3331 & 1.3815 &  & 11.0709 & 8.4265 \\ 
  20 & 0.0144 & 0.0135 &  & 0.0151 & 0.0148 &  & 1.2388 & 1.0030 &  & 7.6560 & 5.9555 \\ 
  100  & 0.0029 & 0.0030 &  & 0.0030 & 0.0030 &  & 2.3889 & 1.6765 &  & 11.2417 & 10.2040 \\ 
  500  & 0.0006 & 0.0006 &  & 0.0006 & 0.0006 &  & 13.7362 & 13.8815 &  & 26.0416 & 25.7790 \\ 
  1,000 & 0.0003 & 0.0003 &  & 0.0003 & 0.0003 &  & 32.9581 & 34.1270 &  & 44.7975 & 44.6580 \\ 
\hline
\multirow{2}{*}{\textbf{marginal}} & \multicolumn{2}{c}{\textbf{500,000}} & & \multicolumn{2}{c}{\textbf{5,000,000}} & & \multicolumn{2}{c}{\textbf{500,000}} & & \multicolumn{2}{c}{\textbf{5,000,000}} \\ 
\cline{2-3} \cline{5-6} \cline{8-9} \cline{11-12} 
&  mean & median & & mean & median & & mean & median & & mean & median \\ \hline
  10 & 0.0499 & 0.0498 &  & 0.0498 & 0.0498 &  & 2.1031 & 2.1010 &  & 26.8011 & 26.8015 \\ 
  20 & 0.0233 & 0.0233 &  & 0.0233 & 0.0233 &  & 3.8852 & 3.8910 &  & 49.4683 & 49.5160 \\ 
  100 & 0.0055 & 0.0055 &  & 0.0055 & 0.0055 &  & 84.2894 & 84.2345 &  & 314.3510 & 314.2770 \\ 
\hline\hline
\end{tabular}
\end{table}

\begin{table}[ht]
\centering
\scriptsize
\caption{This table considers the continuous simulation setting with $\rho=0.5$ and $\nu=0.7416$. The mean and median bias and run time are shown in function of the number clusters for the $k$-means approximation and the number of boundary values for the marginal approximation. Two data set sizes are considered, namely 500,000 and 5,000,000.}
\label{tab: bias-time, rho=0.5 and nu=0.7416}
\begin{tabular}{cccccccccccc}
\hline\hline
 & \multicolumn{5}{c}{bias} & & \multicolumn{5}{c}{run time (s)}\\ 
 \cline{2-6} \cline{8-12} 
\multirow{2}{*}{\textbf{\textit{k}-means}}  & \multicolumn{2}{c}{\textbf{500,000}} & & \multicolumn{2}{c}{\textbf{5,000,000}} & & \multicolumn{2}{c}{\textbf{500,000}} & & \multicolumn{2}{c}{\textbf{5,000,000}} \\ 
\cline{2-3} \cline{5-6} \cline{8-9} \cline{11-12} 
&  mean & median & & mean & median & & mean & median & & mean & median \\ \hline
10 &  0.0215 & 0.0084 &  & 0.0239 & 0.0082 &  & 1.3308 & 1.3740 &  & 11.1616 & 8.5590 \\ 
  20 &0.0175 & 0.0171 &  & 0.0188 & 0.0196 &  & 1.2366 & 0.9975 &  & 7.7186 & 6.0010 \\ 
  100  & 0.0039 & 0.0038 &  & 0.0041 & 0.0040 &  & 2.3745 & 1.6655 &  & 11.2435 & 10.2005 \\ 
  500  & 0.0008 & 0.0008 &  & 0.0008 & 0.0008 &  & 13.5943 & 13.7045 &  & 25.9223 & 25.6600 \\ 
  1,000 & 0.0004 & 0.0004 &  & 0.0004 & 0.0004 &  & 32.4882 & 33.6255 &  & 44.2479 & 44.1040 \\ 
\hline
\multirow{2}{*}{\textbf{marginal}} & \multicolumn{2}{c}{\textbf{500,000}} & & \multicolumn{2}{c}{\textbf{5,000,000}} & & \multicolumn{2}{c}{\textbf{500,000}} & & \multicolumn{2}{c}{\textbf{5,000,000}} \\ 
\cline{2-3} \cline{5-6} \cline{8-9} \cline{11-12} 
&  mean & median & & mean & median & & mean & median & & mean & median \\ \hline
  10 & 0.0492 & 0.0492 &  & 0.0492 & 0.0492 &  & 2.1101 & 2.1120 &  & 26.8251 & 26.7965 \\ 
  20 & 0.0263 & 0.0263 &  & 0.0263 & 0.0263 &  & 3.8448 & 3.8470 &  & 49.1135 & 49.1770 \\ 
  100  & 0.0060 & 0.0060 &  & 0.0060 & 0.0060 &  & 67.8301 & 67.8420 &  & 299.4398 & 299.2035 \\ 
\hline\hline
\end{tabular}
\end{table}

\begin{table}[ht]
\centering
\scriptsize
\caption{This table considers the continuous simulation setting with $\rho=0.75$ and $\nu=0$. The mean and median bias and run time are shown in function of the number clusters for the $k$-means approximation and the number of boundary values for the marginal approximation. Two data set sizes are considered, namely 500,000 and 5,000,000.}
\label{tab: bias-time,  rho=0.75 and nu=0}
\begin{tabular}{cccccccccccc}
\hline\hline
 & \multicolumn{5}{c}{bias} & & \multicolumn{5}{c}{run time (s)}\\ 
 \cline{2-6} \cline{8-12} 
\multirow{2}{*}{\textbf{\textit{k}-means}}  & \multicolumn{2}{c}{\textbf{500,000}} & & \multicolumn{2}{c}{\textbf{5,000,000}} & & \multicolumn{2}{c}{\textbf{500,000}} & & \multicolumn{2}{c}{\textbf{5,000,000}} \\ 
\cline{2-3} \cline{5-6} \cline{8-9} \cline{11-12} 
&  mean & median & & mean & median & & mean & median & & mean & median \\ \hline
10 &  0.0335 & 0.0122 &  & 0.0478 & 0.0416 &  & 1.4215 & 1.4720 &  & 10.6566 & 8.5950 \\ 
  20 & 0.0292 & 0.0279 &  & 0.0319 & 0.0322 &  & 1.6224 & 1.3290 &  & 9.2174 & 8.9905 \\ 
  100  & 0.0066 & 0.0065 &  & 0.0066 & 0.0066 &  & 2.3701 & 1.6160 &  & 11.1091 & 9.9235 \\ 
  500 & 0.0013 & 0.0013 &  & 0.0012 & 0.0012 &  & 12.7239 & 12.2810 &  & 25.7784 & 25.2040 \\ 
  1,000 & 0.0006 & 0.0006 &  & 0.0006 & 0.0006 &  & 31.0945 & 31.4045 &  & 44.3241 & 44.0440 \\ 
\hline
\multirow{2}{*}{\textbf{marginal}} & \multicolumn{2}{c}{\textbf{500,000}} & & \multicolumn{2}{c}{\textbf{5,000,000}} & & \multicolumn{2}{c}{\textbf{500,000}} & & \multicolumn{2}{c}{\textbf{5,000,000}} \\ 
\cline{2-3} \cline{5-6} \cline{8-9} \cline{11-12} 
&  mean & median & & mean & median & & mean & median & & mean & median \\ \hline
    10   & 0.0479 & 0.0479 &  & 0.0479 & 0.0479 &  & 2.0728 & 2.0720 &  & 26.8740 & 26.8965 \\ 
  20  & 0.0253 & 0.0253 &  & 0.0253 & 0.0253 &  & 3.6706 & 3.6660 &  & 49.1315 & 49.1400 \\ 
  100  & 0.0053 & 0.0053 &  & 0.0053 & 0.0053 &  & 18.6860 & 18.6460 &  & 250.4356 & 250.7675 \\ \hline\hline
\end{tabular}
\end{table}

\begin{table}[ht]
\centering
\scriptsize
\caption{This table considers the continuous simulation setting with $\rho=0.75$ and $\nu=0.3583$. The mean and median bias and run time are shown in function of the number clusters for the $k$-means approximation and the number of boundary values for the marginal approximation. Two data set sizes are considered, namely 500,000 and 5,000,000.}
\label{tab: bias-time, rho=0.75 and nu=0.3583}
\begin{tabular}{cccccccccccc}
\hline\hline
 & \multicolumn{5}{c}{bias} & & \multicolumn{5}{c}{run time (s)}\\ 
 \cline{2-6} \cline{8-12} 
\multirow{2}{*}{\textbf{\textit{k}-means}}  & \multicolumn{2}{c}{\textbf{500,000}} & & \multicolumn{2}{c}{\textbf{5,000,000}} & & \multicolumn{2}{c}{\textbf{500,000}} & & \multicolumn{2}{c}{\textbf{5,000,000}} \\ 
\cline{2-3} \cline{5-6} \cline{8-9} \cline{11-12} 
&  mean & median & & mean & median & & mean & median & & mean & median \\ \hline
10 &  0.0519 & 0.0509 &  & 0.0483 & 0.0476 &  & 1.4156 & 1.4635 &  & 10.7063 & 8.6330 \\ 
  20 & 0.0179 & 0.0175 &  & 0.0185 & 0.0179 &  & 1.6191 & 1.3285 &  & 9.2981 & 9.0775 \\ 
  100  & 0.0046 & 0.0045 &  & 0.0043 & 0.0044 &  & 2.3733 & 1.6155 &  & 11.0899 & 9.9065 \\ 
  500  & 0.0008 & 0.0008 &  & 0.0008 & 0.0008 &  & 12.5908 & 12.1410 &  & 25.5900 & 25.0230 \\ 
  1,000  & 0.0004 & 0.0004 &  & 0.0004 & 0.0004 &  & 30.5345 & 30.8385 &  & 43.6687 & 43.3895 \\ 
\hline
\multirow{2}{*}{\textbf{marginal}} & \multicolumn{2}{c}{\textbf{500,000}} & & \multicolumn{2}{c}{\textbf{5,000,000}} & & \multicolumn{2}{c}{\textbf{500,000}} & & \multicolumn{2}{c}{\textbf{5,000,000}} \\ 
\cline{2-3} \cline{5-6} \cline{8-9} \cline{11-12} 
&  mean & median & & mean & median & & mean & median & & mean & median \\ \hline
  10 & 0.0645 & 0.0644 &  & 0.0644 & 0.0644 &  & 2.0915 & 2.0900 &  & 26.9187 & 26.9915 \\ 
  20 & 0.0320 & 0.0320 &  & 0.0320 & 0.0320 &  & 3.8855 & 3.8905 &  & 49.2526 & 49.4965 \\ 
  100 & 0.0080 & 0.0080 &  & 0.0079 & 0.0079 &  & 79.2247 & 79.2360 &  & 314.4613 & 314.6485 \\ 
\hline\hline
\end{tabular}
\end{table}

\begin{table}[ht]
\centering
\scriptsize
\caption{This table considers the continuous simulation setting with $\rho=0.75$ and $\nu=0.7416$. The mean and median bias and run time are shown in function of the number clusters for the $k$-means approximation and the number of boundary values for the marginal approximation. Two data set sizes are considered, namely 500,000 and 5,000,000.}
\label{tab: bias-time, rho=0.75 and nu=0.7416}
\begin{tabular}{cccccccccccc}
\hline\hline
 & \multicolumn{5}{c}{bias} & & \multicolumn{5}{c}{run time (s)}\\ 
 \cline{2-6} \cline{8-12} 
\multirow{2}{*}{\textbf{\textit{k}-means}}  & \multicolumn{2}{c}{\textbf{500,000}} & & \multicolumn{2}{c}{\textbf{5,000,000}} & & \multicolumn{2}{c}{\textbf{500,000}} & & \multicolumn{2}{c}{\textbf{5,000,000}} \\ 
\cline{2-3} \cline{5-6} \cline{8-9} \cline{11-12} 
&  mean & median & & mean & median & & mean & median & & mean & median \\ \hline
10 &  -0.0117 & -0.0270 &  & 0.0104 & 0.0064 &  & 1.4151 & 1.4630 &  & 10.6875 & 8.6155 \\ 
  20 & 0.0320 & 0.0322 &  & 0.0314 & 0.0321 &  & 1.6175 & 1.3260 &  & 9.2665 & 9.0200 \\ 
  100 & 0.0047 & 0.0047 &  & 0.0050 & 0.0050 &  & 2.3635 & 1.6065 &  & 11.0586 & 9.8780 \\ 
  500 & 0.0010 & 0.0010 &  & 0.0011 & 0.0011 &  & 12.4493 & 12.0150 &  & 25.4325 & 24.8580 \\ 
  1,000 & 0.0005 & 0.0005 &  & 0.0005 & 0.0005 &  & 29.9562 & 30.2540 &  & 43.0638 & 42.7810 \\ 
\hline
\multirow{2}{*}{\textbf{marginal}} & \multicolumn{2}{c}{\textbf{500,000}} & & \multicolumn{2}{c}{\textbf{5,000,000}} & & \multicolumn{2}{c}{\textbf{500,000}} & & \multicolumn{2}{c}{\textbf{5,000,000}} \\ 
\cline{2-3} \cline{5-6} \cline{8-9} \cline{11-12} 
&  mean & median & & mean & median & & mean & median & & mean & median \\ \hline
  10 & 0.0512 & 0.0512 &  & 0.0512 & 0.0512 &  & 2.0736 & 2.0785 &  & 26.9187 & 26.9545 \\ 
  20 & 0.0297 & 0.0296 &  & 0.0296 & 0.0296 &  & 3.8399 & 3.8440 &  & 49.2491 & 49.3375 \\ 
  100 & 0.0072 & 0.0072 &  & 0.0073 & 0.0073 &  & 63.6601 & 63.6515 &  & 297.8260 & 298.3230 \\ 
\hline\hline
\end{tabular}
\end{table}

\begin{table}[ht]
\centering
\scriptsize
\caption{This table considers the continuous simulation setting with $\rho=0$ and $\nu=0$. The interquartile range and standard deviation over the estimates and over the run time are shown in function of the number of boundary values for the marginal approximation and the number of clusters for the $k$-means approximation. Two data set sizes are considered, namely 500,000 and 5,000,000.}
\label{tab: sd-IQR, rho=0 and nu=0}
\begin{tabular}{cccccccccccc}
\hline\hline
 & \multicolumn{5}{c}{$\hat{C}$} & & \multicolumn{5}{c}{run time (s)}\\ 
 \cline{2-6} \cline{8-12} 
\multirow{2}{*}{\textbf{\textit{k}-means}}  & \multicolumn{2}{c}{\textbf{500,000}} & & \multicolumn{2}{c}{\textbf{5,000,000}} & & \multicolumn{2}{c}{\textbf{500,000}} & & \multicolumn{2}{c}{\textbf{5,000,000}} \\ 
\cline{2-3} \cline{5-6} \cline{8-9} \cline{11-12} 
& $\sigma$ & IQR & & $\sigma$ & IQR & & $\sigma$ & IQR & & $\sigma$ & IQR \\ \hline
 10   & 0.0351 & 0.0491 &  & 0.0359 & 0.0493 &  & 0.4490 & 0.7553 &  & 2.4559 & 3.6705 \\ 
  20  & 0.0172 & 0.0223 &  & 0.0163 & 0.0215 &  & 0.6323 & 0.8053 &  & 2.4941 & 3.8607 \\ 
 100  & 0.0024 & 0.0032 &  & 0.0021 & 0.0029 &  & 1.4116 & 1.8070 &  & 2.5765 & 0.3045 \\ 
  500  & 0.0006 & 0.0008 &  & 0.0003 & 0.0004 &  & 9.6199 & 18.4325 &  & 2.8644 & 0.3145 \\ 
  1,000 & 0.0005 & 0.0007 &  & 0.0002 & 0.0003 &  & 17.0168 & 29.5770 &  & 2.8465 & 0.2853 \\ 
\hline
\multirow{2}{*}{\textbf{marginal}} & \multicolumn{2}{c}{\textbf{500,000}} & & \multicolumn{2}{c}{\textbf{5,000,000}} & & \multicolumn{2}{c}{\textbf{500,000}} & & \multicolumn{2}{c}{\textbf{5,000,000}} \\ 
\cline{2-3} \cline{5-6} \cline{8-9} \cline{11-12} 
& $\sigma$ & IQR & & $\sigma$ & IQR & & $\sigma$ & IQR & & $\sigma$ & IQR \\ \hline
10 & 0.0006 & 0.0008 &  & 0.0002 & 0.0002 &  & 0.0523 & 0.0702 &  & 1.7858 & 2.5702 \\ 
  20  & 0.0005 & 0.0007 &  & 0.0002 & 0.0002 &  & 0.0425 & 0.0540 &  & 3.0504 & 4.2420 \\ 
  100  & 0.0005 & 0.0006 &  & 0.0002 & 0.0002 &  & 0.1459 & 0.2048 &  & 9.8668 & 14.0612 \\ \hline\hline
\end{tabular}
\end{table}

\begin{table}[ht]
\centering
\scriptsize
\caption{This table considers the continuous simulation setting with $\rho=0$ and $\nu=0.3583$. The interquartile range and standard deviation over the estimates and over the run time are shown in function of the number of boundary values for the marginal approximation and the number of clusters for the $k$-means approximation. Two data set sizes are considered, namely 500,000 and 5,000,000.}
\label{tab: sd-IQR, rho=0 and nu=0.3583$}
\begin{tabular}{cccccccccccc}
\hline\hline
 & \multicolumn{5}{c}{$\hat{C}$} & & \multicolumn{5}{c}{run time (s)}\\ 
 \cline{2-6} \cline{8-12} 
\multirow{2}{*}{\textbf{\textit{k}-means}}  & \multicolumn{2}{c}{\textbf{500,000}} & & \multicolumn{2}{c}{\textbf{5,000,000}} & & \multicolumn{2}{c}{\textbf{500,000}} & & \multicolumn{2}{c}{\textbf{5,000,000}} \\ 
\cline{2-3} \cline{5-6} \cline{8-9} \cline{11-12} 
& $\sigma$ & IQR & & $\sigma$ & IQR & & $\sigma$ & IQR & & $\sigma$ & IQR \\ \hline
10 &  0.0320 & 0.0418 &  & 0.0332 & 0.0416 &  & 0.4474 & 0.7668 &  & 2.4760 & 3.7098 \\ 
  20 & 0.0167 & 0.0221 &  & 0.0161 & 0.0218 &  & 0.6322 & 0.8092 &  & 2.5068 & 3.9133 \\ 
  100 & 0.0026 & 0.0034 &  & 0.0022 & 0.0029 &  & 1.4119 & 1.8040 &  & 2.5677 & 0.3075 \\ 
  500 & 0.0007 & 0.0009 &  & 0.0003 & 0.0004 &  & 9.6294 & 18.4737 &  & 2.8580 & 0.3050 \\ 
  1,000 & 0.0006 & 0.0008 &  & 0.0002 & 0.0003 &  & 17.0606 & 29.6748 &  & 2.8474 & 0.3225 \\ 
\hline
\multirow{2}{*}{\textbf{marginal}} & \multicolumn{2}{c}{\textbf{500,000}} & & \multicolumn{2}{c}{\textbf{5,000,000}} & & \multicolumn{2}{c}{\textbf{500,000}} & & \multicolumn{2}{c}{\textbf{5,000,000}} \\ 
\cline{2-3} \cline{5-6} \cline{8-9} \cline{11-12} 
& $\sigma$ & IQR & & $\sigma$ & IQR & & $\sigma$ & IQR & & $\sigma$ & IQR \\ \hline
  10 & 0.0007 & 0.0010 &  & 0.0002 & 0.0003 &  & 0.0507 & 0.0772 &  & 1.8203 & 2.7230 \\ 
  20 & 0.0007 & 0.0009 &  & 0.0002 & 0.0003 &  & 0.0497 & 0.0650 &  & 3.0622 & 4.1202 \\ 
  100 & 0.0006 & 0.0008 &  & 0.0002 & 0.0002 &  & 0.6802 & 1.0220 &  & 9.8971 & 16.1810 \\ 
\hline\hline
\end{tabular}
\end{table}

\begin{table}[ht]
\centering
\scriptsize
\caption{This table considers the continuous simulation setting with $\rho=0$ and $\nu=0.7416$. The interquartile range and standard deviation over the estimates and over the run time are shown in function of the number of boundary values for the marginal approximation and the number of clusters for the $k$-means approximation. Two data set sizes are considered, namely 500,000 and 5,000,000.}
\label{tab: sd-IQR, rho=0 and nu=0.7416}
\begin{tabular}{cccccccccccc}
\hline\hline
 & \multicolumn{5}{c}{$\hat{C}$} & & \multicolumn{5}{c}{run time (s)}\\ 
 \cline{2-6} \cline{8-12} 
\multirow{2}{*}{\textbf{\textit{k}-means}}  & \multicolumn{2}{c}{\textbf{500,000}} & & \multicolumn{2}{c}{\textbf{5,000,000}} & & \multicolumn{2}{c}{\textbf{500,000}} & & \multicolumn{2}{c}{\textbf{5,000,000}} \\ 
\cline{2-3} \cline{5-6} \cline{8-9} \cline{11-12} 
& $\sigma$ & IQR & & $\sigma$ & IQR & & $\sigma$ & IQR & & $\sigma$ & IQR \\ \hline
10 &  0.0423 & 0.0587 &  & 0.0439 & 0.0586 &  & 0.4471 & 0.7527 &  & 2.4578 & 3.6938 \\ 
  20 & 0.0254 & 0.0351 &  & 0.0223 & 0.0306 &  & 0.6308 & 0.8033 &  & 2.5044 & 3.8827 \\ 
  100 & 0.0028 & 0.0035 &  & 0.0025 & 0.0033 &  & 1.4142 & 1.8052 &  & 2.5722 & 0.2963 \\ 
  500 & 0.0008 & 0.0011 &  & 0.0004 & 0.0005 &  & 9.6469 & 18.4815 &  & 2.8590 & 0.3043 \\ 
  1,000 & 0.0007 & 0.0009 &  & 0.0003 & 0.0003 &  & 17.0871 & 29.7192 &  & 2.8452 & 0.3060 \\ 
\hline
\multirow{2}{*}{\textbf{marginal}} & \multicolumn{2}{c}{\textbf{500,000}} & & \multicolumn{2}{c}{\textbf{5,000,000}} & & \multicolumn{2}{c}{\textbf{500,000}} & & \multicolumn{2}{c}{\textbf{5,000,000}} \\ 
\cline{2-3} \cline{5-6} \cline{8-9} \cline{11-12} 
& $\sigma$ & IQR & & $\sigma$ & IQR & & $\sigma$ & IQR & & $\sigma$ & IQR \\ \hline
  10 & 0.0009 & 0.0012 &  & 0.0003 & 0.0004 &  & 0.0561 & 0.0903 &  & 1.7996 & 2.4957 \\ 
  20 & 0.0008 & 0.0010 &  & 0.0002 & 0.0003 &  & 0.0800 & 0.1218 &  & 2.9585 & 4.0628 \\ 
  100 & 0.0007 & 0.0009 &  & 0.0002 & 0.0003 &  & 0.5187 & 0.7805 &  & 10.0947 & 16.7948 \\ 
\hline\hline
\end{tabular}
\end{table}

\begin{table}[ht]
\centering
\scriptsize
\caption{This table considers the continuous simulation setting with $\rho=0.25$ and $\nu=0$. The interquartile range and standard deviation over the estimates and over the run time are shown in function of the number of boundary values for the marginal approximation and the number of clusters for the $k$-means approximation. Two data set sizes are considered, namely 500,000 and 5,000,000.}
\label{tab: sd-IQR, rho=0.25 and nu=0}
\begin{tabular}{cccccccccccc}
\hline\hline
 & \multicolumn{5}{c}{$\hat{C}$} & & \multicolumn{5}{c}{run time (s)}\\ 
 \cline{2-6} \cline{8-12} 
\multirow{2}{*}{\textbf{\textit{k}-means}}  & \multicolumn{2}{c}{\textbf{500,000}} & & \multicolumn{2}{c}{\textbf{5,000,000}} & & \multicolumn{2}{c}{\textbf{500,000}} & & \multicolumn{2}{c}{\textbf{5,000,000}} \\ 
\cline{2-3} \cline{5-6} \cline{8-9} \cline{11-12} 
& $\sigma$ & IQR & & $\sigma$ & IQR & & $\sigma$ & IQR & & $\sigma$ & IQR \\ \hline
10 &  0.0378 & 0.0468 &  & 0.0347 & 0.0427 &  & 0.5111 & 0.6972 &  & 2.3339 & 3.6152 \\ 
  20 & 0.0179 & 0.0248 &  & 0.0163 & 0.0211 &  & 0.6767 & 0.8275 &  & 2.4550 & 3.9225 \\ 
  100 & 0.0025 & 0.0031 &  & 0.0020 & 0.0027 &  & 1.4298 & 1.8153 &  & 2.7002 & 0.3020 \\ 
  500 & 0.0006 & 0.0008 &  & 0.0003 & 0.0004 &  & 9.3448 & 18.3017 &  & 2.5973 & 0.3035 \\ 
  1,000 & 0.0005 & 0.0006 &  & 0.0002 & 0.0003 &  & 16.8166 & 29.3665 &  & 2.8414 & 0.3348 \\ 
\hline
\multirow{2}{*}{\textbf{marginal}} & \multicolumn{2}{c}{\textbf{500,000}} & & \multicolumn{2}{c}{\textbf{5,000,000}} & & \multicolumn{2}{c}{\textbf{500,000}} & & \multicolumn{2}{c}{\textbf{5,000,000}} \\ 
\cline{2-3} \cline{5-6} \cline{8-9} \cline{11-12} 
& $\sigma$ & IQR & & $\sigma$ & IQR & & $\sigma$ & IQR & & $\sigma$ & IQR \\ \hline
  10  & 0.0005 & 0.0007 &  & 0.0002 & 0.0002 &  & 0.0522 & 0.0690 &  & 1.8174 & 2.5452 \\ 
  20  & 0.0005 & 0.0006 &  & 0.0002 & 0.0002 &  & 0.0397 & 0.0540 &  & 3.1205 & 4.2248 \\ 
  100  & 0.0004 & 0.0006 &  & 0.0001 & 0.0002 &  & 0.2279 & 0.3653 &  & 8.8337 & 11.3668 \\  
\hline\hline
\end{tabular}
\end{table}

\begin{table}[ht]
\centering
\scriptsize
\caption{This table considers the continuous simulation setting with $\rho=0.25$ and $\nu=0.7416$. The interquartile range and standard deviation over the estimates and over the run time are shown in function of the number of boundary values for the marginal approximation and the number of clusters for the $k$-means approximation. Two data set sizes are considered, namely 500,000 and 5,000,000.}
\label{tab: sd-IQR, rho=0.25 and nu=0.7416}
\begin{tabular}{cccccccccccc}
\hline\hline
 & \multicolumn{5}{c}{$\hat{C}$} & & \multicolumn{5}{c}{run time (s)}\\ 
 \cline{2-6} \cline{8-12} 
\multirow{2}{*}{\textbf{\textit{k}-means}}  & \multicolumn{2}{c}{\textbf{500,000}} & & \multicolumn{2}{c}{\textbf{5,000,000}} & & \multicolumn{2}{c}{\textbf{500,000}} & & \multicolumn{2}{c}{\textbf{5,000,000}} \\ 
\cline{2-3} \cline{5-6} \cline{8-9} \cline{11-12} 
& $\sigma$ & IQR & & $\sigma$ & IQR & & $\sigma$ & IQR & & $\sigma$ & IQR \\ \hline
10 &  0.0443 & 0.0607 &  & 0.0452 & 0.0599 &  & 0.5092 & 0.6895 &  & 2.3034 & 3.5803 \\ 
  20 & 0.0262 & 0.0366 &  & 0.0222 & 0.0300 &  & 0.6744 & 0.8155 &  & 2.4331 & 3.9035 \\ 
  100 & 0.0027 & 0.0034 &  & 0.0025 & 0.0033 &  & 1.4280 & 1.8105 &  & 2.6872 & 0.3048 \\ 
  500 & 0.0007 & 0.0009 &  & 0.0004 & 0.0005 &  & 9.4002 & 18.3627 &  & 2.5976 & 0.2850 \\ 
  1,000 & 0.0006 & 0.0009 &  & 0.0002 & 0.0003 &  & 16.8475 & 29.5138 &  & 2.8394 & 0.3052 \\
\hline
\multirow{2}{*}{\textbf{marginal}} & \multicolumn{2}{c}{\textbf{500,000}} & & \multicolumn{2}{c}{\textbf{5,000,000}} & & \multicolumn{2}{c}{\textbf{500,000}} & & \multicolumn{2}{c}{\textbf{5,000,000}} \\ 
\cline{2-3} \cline{5-6} \cline{8-9} \cline{11-12} 
& $\sigma$ & IQR & & $\sigma$ & IQR & & $\sigma$ & IQR & & $\sigma$ & IQR \\ \hline 
  10 & 0.0008 & 0.0011 &  & 0.0003 & 0.0003 &  & 0.0545 & 0.0880 &  & 1.8056 & 2.4965 \\ 
  20 & 0.0007 & 0.0009 &  & 0.0002 & 0.0003 &  & 0.0688 & 0.0940 &  & 3.0020 & 4.1795 \\ 
  100 & 0.0006 & 0.0008 &  & 0.0002 & 0.0003 &  & 0.5745 & 0.8470 &  & 9.2107 & 12.8475 \\ 
\hline\hline
\end{tabular}
\end{table}

\begin{table}[ht]
\centering
\scriptsize
\caption{This table considers the continuous simulation setting with $\rho=0.5$ and $\nu=0$. The interquartile range and standard deviation over the estimates and over the run time are shown in function of the number of boundary values for the marginal approximation and the number of clusters for the $k$-means approximation. Two data set sizes are considered, namely 500,000 and 5,000,000.}
\label{tab: sd-IQR, rho=0.5 and nu=0}
\begin{tabular}{cccccccccccc}
\hline\hline
 & \multicolumn{5}{c}{$\hat{C}$} & & \multicolumn{5}{c}{run time (s)}\\ 
 \cline{2-6} \cline{8-12} 
\multirow{2}{*}{\textbf{\textit{k}-means}}  & \multicolumn{2}{c}{\textbf{500,000}} & & \multicolumn{2}{c}{\textbf{5,000,000}} & & \multicolumn{2}{c}{\textbf{500,000}} & & \multicolumn{2}{c}{\textbf{5,000,000}} \\ 
\cline{2-3} \cline{5-6} \cline{8-9} \cline{11-12} 
& $\sigma$ & IQR & & $\sigma$ & IQR & & $\sigma$ & IQR & & $\sigma$ & IQR \\ \hline
10 &  0.0304 & 0.0114 &  & 0.0310 & 0.0274 &  & 0.5886 & 0.9725 &  & 7.0572 & 11.9343 \\ 
  20 & 0.0161 & 0.0222 &  & 0.0150 & 0.0205 &  & 0.7020 & 0.9138 &  & 2.8837 & 4.0043 \\ 
  100 & 0.0025 & 0.0030 &  & 0.0020 & 0.0027 &  & 1.5403 & 2.1868 &  & 2.6976 & 0.2988 \\ 
  500 & 0.0005 & 0.0007 &  & 0.0003 & 0.0004 &  & 9.0546 & 16.7807 &  & 2.2227 & 0.3317 \\ 
  1,000 & 0.0004 & 0.0005 &  & 0.0002 & 0.0002 &  & 16.3951 & 29.0728 &  & 1.9932 & 0.2795 \\ 
\hline
\multirow{2}{*}{\textbf{marginal}} & \multicolumn{2}{c}{\textbf{500,000}} & & \multicolumn{2}{c}{\textbf{5,000,000}} & & \multicolumn{2}{c}{\textbf{500,000}} & & \multicolumn{2}{c}{\textbf{5,000,000}} \\ 
\cline{2-3} \cline{5-6} \cline{8-9} \cline{11-12} 
& $\sigma$ & IQR & & $\sigma$ & IQR & & $\sigma$ & IQR & & $\sigma$ & IQR \\ \hline
  10  & 0.0005 & 0.0006 &  & 0.0002 & 0.0002 &  & 0.0702 & 0.0872 &  & 1.7800 & 2.6195 \\ 
  20  & 0.0004 & 0.0006 &  & 0.0001 & 0.0002 &  & 0.0845 & 0.1150 &  & 3.9106 & 4.5258 \\ 
  100  & 0.0004 & 0.0005 &  & 0.0001 & 0.0002 &  & 0.2957 & 0.4195 &  & 7.8842 & 10.0610 \\  
\hline\hline
\end{tabular}
\end{table}

\begin{table}[ht]
\centering
\scriptsize
\caption{This table considers the continuous simulation setting with $\rho=0.5$ and $\nu=0.3583$. The interquartile range and standard deviation over the estimates and over the run time are shown in function of the number of boundary values for the marginal approximation and the number of clusters for the $k$-means approximation. Two data set sizes are considered, namely 500,000 and 5,000,000.}
\label{tab: sd-IQR, rho=0.5 and nu=0.3583}
\begin{tabular}{cccccccccccc}
\hline\hline
 & \multicolumn{5}{c}{$\hat{C}$} & & \multicolumn{5}{c}{run time (s)}\\ 
 \cline{2-6} \cline{8-12} 
\multirow{2}{*}{\textbf{\textit{k}-means}}  & \multicolumn{2}{c}{\textbf{500,000}} & & \multicolumn{2}{c}{\textbf{5,000,000}} & & \multicolumn{2}{c}{\textbf{500,000}} & & \multicolumn{2}{c}{\textbf{5,000,000}} \\ 
\cline{2-3} \cline{5-6} \cline{8-9} \cline{11-12} 
& $\sigma$ & IQR & & $\sigma$ & IQR & & $\sigma$ & IQR & & $\sigma$ & IQR \\ \hline
10 &  0.0291 & 0.0383 &  & 0.0302 & 0.0496 &  & 0.5902 & 0.9797 &  & 7.0510 & 11.8870 \\ 
  20 & 0.0160 & 0.0225 &  & 0.0160 & 0.0209 &  & 0.7057 & 0.9110 &  & 2.8767 & 3.9880 \\ 
  100 & 0.0025 & 0.0030 &  & 0.0021 & 0.0028 &  & 1.5494 & 2.2068 &  & 2.7107 & 0.3110 \\ 
  500 & 0.0006 & 0.0007 &  & 0.0003 & 0.0004 &  & 9.0493 & 16.7798 &  & 2.2135 & 0.3003 \\ 
  1,000 & 0.0005 & 0.0006 &  & 0.0002 & 0.0002 &  & 16.4039 & 29.2357 &  & 2.0005 & 0.2762 \\ 
\hline
\multirow{2}{*}{\textbf{marginal}} & \multicolumn{2}{c}{\textbf{500,000}} & & \multicolumn{2}{c}{\textbf{5,000,000}} & & \multicolumn{2}{c}{\textbf{500,000}} & & \multicolumn{2}{c}{\textbf{5,000,000}} \\ 
\cline{2-3} \cline{5-6} \cline{8-9} \cline{11-12} 
& $\sigma$ & IQR & & $\sigma$ & IQR & & $\sigma$ & IQR & & $\sigma$ & IQR \\ \hline
  10 & 0.0008 & 0.0008 &  & 0.0002 & 0.0003 &  & 0.0520 & 0.0850 &  & 1.8610 & 2.7732 \\ 
  20 & 0.0005 & 0.0007 &  & 0.0002 & 0.0002 &  & 0.0678 & 0.0783 &  & 3.1378 & 4.2897 \\ 
  100 & 0.0005 & 0.0006 &  & 0.0002 & 0.0002 &  & 0.7482 & 0.9652 &  & 8.2329 & 10.1155 \\ 
\hline\hline
\end{tabular}
\end{table}

\begin{table}[ht]
\centering
\scriptsize
\caption{This table considers the continuous simulation setting with $\rho=0.5$ and $\nu=0.7416$. The interquartile range and standard deviation over the estimates and over the run time are shown in function of the number of boundary values for the marginal approximation and the number of clusters for the $k$-means approximation. Two data set sizes are considered, namely 500,000 and 5,000,000.}
\label{tab: sd-IQR, rho=0.5 and nu=0.7416}
%
\end{table}

\begin{table}[ht]
\centering
\scriptsize
\caption{This table considers the continuous simulation setting with $\rho=0.75$ and $\nu=0$. The interquartile range and standard deviation over the estimates and over the run time are shown in function of the number of boundary values for the marginal approximation and the number of clusters for the $k$-means approximation. Two data set sizes are considered, namely 500,000 and 5,000,000.}
\label{tab: sd-IQR, rho=0.75 and nu=0}
%
\end{table}
\FloatBarrier
\vskip 0.2in

\bibliographystyle{plainnat}
\bibliography{bib}

\begin{thebibliography}{18}
\providecommand{\natexlab}[1]{#1}
\providecommand{\url}[1]{\texttt{#1}}
\expandafter\ifx\csname urlstyle\endcsname\relax
  \providecommand{\doi}[1]{doi: #1}\else
  \providecommand{\doi}{doi: \begingroup \urlstyle{rm}\Url}\fi

\bibitem[Animesh(2019)]{CTR}
Animesh.
\newblock Predict click through rate (ctr) for a website, 2019.
\newblock URL
  \url{https://www.kaggle.com/animeshgoyal9/predict-click-through-rate-ctr-for-a-website}.

\bibitem[Bamber(1975)]{bamber1975area}
Donald Bamber.
\newblock The area above the ordinal dominance graph and the area below the
  receiver operating characteristic graph.
\newblock \emph{Journal of mathematical psychology}, 12\penalty0 (4):\penalty0
  387--415, 1975.

\bibitem[Bouckaert(2006)]{bouckaert2006efficient}
Remco~R Bouckaert.
\newblock Efficient auc learning curve calculation.
\newblock In \emph{Australasian Joint Conference on Artificial Intelligence},
  pages 181--191. Springer, 2006.

\bibitem[Calders and Jaroszewicz(2007)]{calders2007efficient}
Toon Calders and Szymon Jaroszewicz.
\newblock Efficient auc optimization for classification.
\newblock In \emph{European Conference on Principles of Data Mining and
  Knowledge Discovery}, pages 42--53. Springer, 2007.

\bibitem[Eguchi and Copas(2002)]{eguchi2002class}
Shinto Eguchi and John Copas.
\newblock A class of logistic-type discriminant functions.
\newblock \emph{Biometrika}, 89\penalty0 (1):\penalty0 1--22, 2002.

\bibitem[Fawcett(2006)]{fawcett2006introduction}
Tom Fawcett.
\newblock An introduction to roc analysis.
\newblock \emph{Pattern recognition letters}, 27\penalty0 (8):\penalty0
  861--874, 2006.

\bibitem[Gerds and Schumacher(2007)]{Gerds2007}
Thomas~A Gerds and Michael Schumacher.
\newblock Efron-type measures of prediction error for survival analysis.
\newblock \emph{Biometrics}, 63:\penalty0 1283--1287, 2007.

\bibitem[GoogleCloud(2018)]{taxi}
GoogleCloud.
\newblock New york city taxi fare prediction, 2018.
\newblock URL
  \url{https://www.kaggle.com/c/new-york-city-taxi-fare-prediction}.

\bibitem[Harrell et~al.(1982)Harrell, Califf, Pryor, Lee, and
  Rosati]{Harrell1982}
Frank~E Harrell, Robert~M Califf, David~B Pryor, Kerry~L Lee, and Robert~A
  Rosati.
\newblock Evaluating the yield of medical tests.
\newblock \emph{Journal of the American Medical Association}, 247:\penalty0
  2543–2546, 1982.

\bibitem[Heller and Mo(2016)]{heller2016estimating}
Glenn Heller and Qianxing Mo.
\newblock Estimating the concordance probability in a survival analysis with a
  discrete number of risk groups.
\newblock \emph{Lifetime data analysis}, 22\penalty0 (2):\penalty0 263--279,
  2016.

\bibitem[Komori(2011)]{komori2011boosting}
Osamu Komori.
\newblock A boosting method for maximization of the area under the roc curve.
\newblock \emph{Annals of the Institute of Statistical Mathematics},
  63\penalty0 (5):\penalty0 961--979, 2011.

\bibitem[Liu et~al.(2008)Liu, Wu, and Zhou]{liu2008exploratory}
Xu-Ying Liu, Jianxin Wu, and Zhi-Hua Zhou.
\newblock Exploratory undersampling for class-imbalance learning.
\newblock \emph{IEEE Transactions on Systems, Man, and Cybernetics, Part B
  (Cybernetics)}, 39\penalty0 (2):\penalty0 539--550, 2008.

\bibitem[Ma and Huang(2005)]{ma2005regularized}
Shuangge Ma and Jian Huang.
\newblock Regularized roc method for disease classification and biomarker
  selection with microarray data.
\newblock \emph{Bioinformatics}, 21\penalty0 (24):\penalty0 4356--4362, 2005.

\bibitem[Pencina and D'Agostino(2004)]{pencina2004overall}
Michael~J Pencina and Ralph~B D'Agostino.
\newblock Overall c as a measure of discrimination in survival analysis: model
  specific population value and confidence interval estimation.
\newblock \emph{Statistics in medicine}, 23\penalty0 (13):\penalty0 2109--2123,
  2004.

\bibitem[Reddy and Aggarwal(2015)]{reddy2015healthcare}
Chandan~K Reddy and Charu~C Aggarwal.
\newblock \emph{Healthcare data analytics}, volume~36.
\newblock CRC Press, 2015.

\bibitem[Steyerberg et~al.(2010)Steyerberg, Vickers, Cook, Gerds, Gonen,
  Obuchowski, Pencina, and Kattan]{steyerberg2010assessing}
Ewout~W Steyerberg, Andrew~J Vickers, Nancy~R Cook, Thomas Gerds, Mithat Gonen,
  Nancy Obuchowski, Michael~J Pencina, and Michael~W Kattan.
\newblock Assessing the performance of prediction models: a framework for some
  traditional and novel measures.
\newblock \emph{Epidemiology (Cambridge, Mass.)}, 21\penalty0 (1):\penalty0
  128, 2010.

\bibitem[van Breemen(2018)]{taxiNotebook}
Albert van Breemen.
\newblock Nyc taxi fare - data exploration, 2018.
\newblock URL
  \url{https://www.kaggle.com/breemen/nyc-taxi-fare-data-exploration}.

\bibitem[Yan and Greene(2008)]{yan2008investigating}
Guofen Yan and Tom Greene.
\newblock Investigating the effects of ties on measures of concordance.
\newblock \emph{Statistics in medicine}, 27\penalty0 (21):\penalty0 4190--4206,
  2008.

\end{thebibliography}
\end{document}